\documentclass[11pt, reqno]{amsart}
\usepackage{amsfonts,latexsym, enumerate}
\usepackage{amsmath}
\usepackage{amscd}
\usepackage{float,amsmath,amssymb,mathrsfs,bm,multirow,graphics}
\usepackage[dvips]{graphicx}
\usepackage[percent]{overpic}
\usepackage[pdftex]{color}
\usepackage{amsaddr}
\usepackage[numbers,sort&compress]{natbib}

\newcommand{\nequation}{\setcounter{equation}{0}}
\renewcommand{\theequation}{\mbox{\arabic{section}.\arabic{equation}}}
\newcommand{\R}{{\Bbb R}}

\newcommand{\C}{{\Bbb C}}

\newcommand{\D}{{\Bbb D}}

\newcommand{\proofbegin}{\noindent{\it Proof.\quad}}
\newcommand{\proofend}{\hfill$\Box$\bigskip}
\newcommand{\proofendcontinue}{\hfill \raisebox{.8mm}[0cm][0cm]{$\bigtriangledown$}\bigskip}

\newcommand{\re}{\text{\upshape Re\,}}
\newcommand{\Arg}{\text{\upshape Arg\,}}
\newcommand{\im}{\text{\upshape Im\,}}

\newcommand{\ntlim}{\lim^\angle}

\newtheorem{theorem}{Theorem}[section]

\newtheorem{lemma}[theorem]{Lemma}

\newtheorem{remark}[theorem]{Remark}

\newtheorem{figuretext}{Figure}

\input epsf
\title[The nonlinear steepest descent method]
{The nonlinear steepest descent method for Riemann-Hilbert problems of low regularity}

\author{Jonatan Lenells}
\address{Department of Mathematics, KTH Royal Institute of Technology, \\ 100 44 Stockholm, Sweden.}
\email{jlenells@kth.se}

\begin{document}

\begin{abstract} 
\noindent
We prove a nonlinear steepest descent theorem for Riemann-Hilbert problems with Carleson jump contours and jump matrices of low regularity and slow decay. We illustrate the theorem by deriving the long-time asymptotics for the mKdV equation in the similarity sector for initial data with limited decay and regularity. 
\end{abstract}

\maketitle



\setcounter{tocdepth}{1}
\tableofcontents

\section{Introduction}\nequation
The method of nonlinear steepest descent was introduced in the early 1990's in a seminal paper by Deift and Zhou \cite{DZ1993}, building on earlier work of Manakov \cite{M1974} and Its \cite{I1981}. Whereas the classical steepest descent method yields asymptotic expansions as $t \to \infty$ of scalar integrals of the form
\begin{align}\label{basicintegral}
I(t) = \int_\Gamma f(z) e^{t\Phi(z)} dz
\end{align}
where $\Gamma$ is a contour in the complex $z$-plane, the nonlinear steepest descent method yields expansions of solutions of matrix Riemann-Hilbert (RH) problems. In the same way that the solutions of a large class of problems involving linear differential equations can be represented by scalar integrals of the form (\ref{basicintegral}), the solutions of many nonlinear problems can be represented in terms of solutions of matrix RH problems. 

The nonlinear steepest descent method relies on the same idea as its classical analog. The jump matrix of the RH problem contains basic exponentials of the form $e^{\pm t\Phi(z)}$. By deforming the contour so that the jump involves only $e^{t\Phi(z)}$ when $z$ lies in the domain $\{\re \Phi(z) < 0\}$ whereas it involves only $e^{-t\Phi(z)}$ when $z$ lies in the domain $\{\re \Phi(z) > 0\}$, it is ensured that the jump is small for large $t$ except near a small number of `critical points' at which $\re \Phi(z) = 0$. Near each critical point the RH problem converges as $t \to \infty$ to a RH problem which can be explicitly solved. Hence, as in the classical method, an asymptotic expansion of the solution can be obtained by summing up the contributions from the individual critical points. The method can be generalized to also allow for jumps that are small for all large $t$ except near a number of `critical bands' \cite{DVZ1997}.

The purpose of this paper is to implement the nonlinear steepest descent method for RH problems of low regularity. More precisely, we prove a nonlinear steepest descent theorem (see Theorem \ref{steepestdescentth}) applicable to Riemann-Hilbert problems with Carleson jump contours and jump matrices of low regularity and slow decay (both as $t \to \infty$ and as $z \to \infty$). We illustrate the theorem by deriving the long-time asymptotics for the mKdV equation 
\begin{align}\label{mkdv}
u_t - 6 u^2u_x + u_{xxx} = 0, \qquad x \in \R, \  t \geq 0,
\end{align} 
in the similarity sector for initial data with limited decay and regularity (see Theorem \ref{mainth1}). 

In \cite{DZ1993}, formulas were established for the asymptotics of the solution of (\ref{mkdv})
under the assumption that the initial data $u_0(x) = u(x, 0)$ lie in the Schwartz class $\mathcal{S}(\R)$ of rapidly decreasing functions. 
Even though the main idea of our approach is the same as in \cite{DZ1993}, the proof proceeds along somewhat different lines: By isolating the dominant contributions of the critical points directly in an appropriately rescaled RH problem, we can easily find the asymptotics using Cauchy's formula. In this way, we avoid having to establish a number of operator identities related to the restriction and decoupling of various parts of the jump contour. 
Our goal is to provide a rigorous and concise treatment; precise and uniform error estimates are presented throughout the paper. 

Our approach relies heavily on the fact that the theory of $L^2$-RH problems can be generalized to the case of sectionally analytic functions with jumps across Carleson contours. The formulation of a successful Riemann-Hilbert theory is intricately linked to the boundedness of the Cauchy singular operator $\mathcal{S}_\Gamma$. Following a long development involving many researchers, it became clear in the 1980s that $\mathcal{S}_\Gamma$ is bounded on the weighted Lebesgue space $L^p(\Gamma, w)$, $1 < p < \infty$, if and only if $\Gamma$ is a Carleson curve and $w$ is a Muckenhoupt weight, see \cite{BK1997}.  This result makes it possible to develop a theory for $L^p$-RH problems with Carleson contours (see \cite{LCarleson} and references therein). In this theory, (generalized) Smirnoff classes of analytic functions play an important role and make it possible to formulate precise conditions under which, for example, contour deformations can take place. Here, by taking advantage of these techniques, we are able to present the nonlinear steepest descent method in a very general setting. 

Our Theorem \ref{steepestdescentth} generalizes the nonlinear steepest descent argument of \cite{DZ1993} in two ways: 
\begin{itemize}
\item[$(a)$] Theorem \ref{steepestdescentth} allows for very general contours. Whereas the arguments of \cite{DZ1993} were adapted to the case of piecewise smooth contours, Theorem 2.1 applies in the context of a general Carleson contour. This generalization is not needed for the application to mKdV on the line, but is important for other applications. 

\item[$(b)$] Theorem \ref{steepestdescentth} allows for jump matrices with low regularity and slow decay. In the application to the mKdV equation presented in Section \ref{similaritysec}, we utilize this fact to find a formula for the asymptotics of the solution $u(x,t)$ for initial data $u_0$ in $C^5(\R)$ under the decay assumption $(1+|x|)^{11}u_0^{(i)}(x) \in L^1(\R)$ for $i = 0,1, \dots, 5$. In \cite{DZ1993} it was assumed that $u_0$ belongs to the Schwartz class $\mathcal{S}(\R)$. 
\end{itemize}

Certain steps in the proof of Theorem \ref{steepestdescentth} are inspired by the nice presentation of \cite{GT2009} (see also \cite{KT2009}). In particular, our definition of $\hat{v}$ (see equation (\ref{hatvdef})) is similar to equation (5.13) of \cite{GT2009}. However, whereas \cite{GT2009} utilizes a solution $M_j$ of a RH problem obtained by restricting the jump to a small cross in an $\epsilon$ neighborhood of the critical point, we instead compare the RH solution directly to the solution of the model RH problem. This leads to a more straightforward presentation and circumvents some implicit difficulties in \cite{GT2009} related to the fact that $M_j$ in general has singularities at the endpoints of the small cross. 

The class of Carleson contours is very large; for example, it includes contours with cusps and nontransversal intersections. This means that our approach can be used to rigorously analyze asymptotics for a large class of RH problems. RH problems with contours involving nontransversal intersections are important in the study of initial-boundary problems for integrable evolution equations, see e.g. \cite{L2013}. In this paper, we consider equation (\ref{mkdv}) on the line, but more complicated examples for initial and initial-boundary value problems will be analyzed elsewhere. 

The defocusing mKdV equation (\ref{mkdv}) on the line is locally well-posed in $H^s$ for $s \geq 1/4$ \cite{KPV1993} and globally well-posed in $H^s$ for $s > 1/4$ \cite{CKSTT2003}.
Riemann-Hilbert problems with jump matrices of low regularity are analyzed in \cite{DZ2003}. For problems where the jump matrix fails to be sectionally analytic, a dbar generalization of the nonlinear steepest descent method has been introduced in \cite{DM2008, MM2006} (see also \cite{CJ2014}).

In Section \ref{steepsec}, we prove a nonlinear steepest descent theorem for RH problems with Carleson jump contours. 
In Section \ref{mkdvsec}, we recall how the solution of the mKdV equation can be expressed in terms of the solution of a RH problem. 
In Section \ref{similaritysec}, we derive the long time behavior of (\ref{mkdv}) in the similarity sector for a large class of initial data. 
Some results on $L^2$-RH problems are collected in Appendix \ref{RHapp}. 
Appendix \ref{exactapp} contains a derivation of the exact solution of the model RH problem which is relevant near the critical points.

\section{A nonlinear steepest descent theorem}\nequation\label{steepsec}
Our first theorem (Theorem \ref{steepestdescentth}) provides an implementation of the nonlinear steepest descent method for RH problems with Carleson jump contours and jump matrices of low regularity and slow decay.\footnote{We refer to Appendix \ref{RHapp} for precise definitions of Carleson jump contours and $L^2$-RH problems.} Although the theorem is formulated, for definiteness, under the assumption that there are two critical points related by reflection in the imaginary axis (this is the situation relevant for the similarity sector of the mKdV equation), it can be readily generalized to scenarios with multiple critical points and/or different symmetries. 

\begin{figure}
\begin{center}
 \begin{overpic}[width=.35\textwidth]{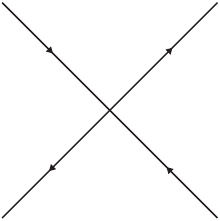}
 \put(67,81){$X_1$}
 \put(24,81){$X_2$}
 \put(24,16){$X_3$}
 \put(67,16){$X_4$}
 \end{overpic}
   \bigskip
   \begin{figuretext}\label{X.pdf}
      The contour $X = X_1 \cup \cdots \cup X_4$.
      \end{figuretext}
   \end{center}
\end{figure}

Let $X$ denote the cross $X = X_1 \cup \cdots \cup X_4 \subset \C$, where the rays
\begin{align} \nonumber
&X_1 = \bigl\{se^{\frac{i\pi}{4}}\, \big| \, 0 \leq s < \infty\bigr\}, && 
X_2 = \bigl\{se^{\frac{3i\pi}{4}}\, \big| \, 0 \leq s < \infty\bigr\},  
	\\ \label{Xdef}
&X_3 = \bigl\{se^{-\frac{3i\pi}{4}}\, \big| \, 0 \leq s < \infty\bigr\}, && 
X_4 = \bigl\{se^{-\frac{i\pi}{4}}\, \big| \, 0 \leq s < \infty\bigr\},
\end{align}
are oriented as in Figure \ref{X.pdf}. 
For $r > 0$, we let $X^r = X_1^r \cup \cdots \cup X_4^r$ denote the restriction of $X$ to the disk of radius $r$ centered at the origin, i.e. $X^r = X \cap \{|z| < r\}$. 
Given a Carleson jump contour $\Gamma$ and $a,b \in \R$ with $a < b$, we call $W_{a,b} = \{a \leq \arg k \leq b\}$ a nontangential sector at $\infty$ if there exists a $\delta > 0$ such that $W_{a - \delta, b + \delta}$ does not intersect $\Gamma \cap \{|z| > R\}$ whenever $R>0$ is large enough. If $f(k)$ is a function of $k \in \C \setminus \Gamma$, we say that $f$ has nontangential limit $L$ at $\infty$, written 
$$\ntlim_{k\to\infty} f(k) = L,$$ 
if $\lim_{\substack{k \to \infty \\ k \in W_{a,b}}}f(k) = L$ for every nontangential sector $W_{a,b}$ at $\infty$. 
Throughout the paper, complex powers and logarithms are defined using the principal branch: If $z, a \in \C$ and $z \neq 0$, then $\ln z := \ln|z| + i\Arg{z}$ and $z^a := e^{a\ln z}$, where $\Arg z \in (-\pi, \pi]$ denotes the principal value of $\arg z$. We use $C$ to denote a generic constant that can change within a computation. The Riemann sphere is denoted by $\hat{\C} = \C \cup \{\infty\}$.
The Smirnoff spaces $\dot{E}^p(D)$ are defined in Appendix \ref{RHapp}.

\begin{theorem}[Nonlinear steepest descent]\label{steepestdescentth}
Let $\mathcal{I} \subset \R$ be a (possibly infinite) interval. Let $\rho, \epsilon, k_0:\mathcal{I} \to (0,\infty)$ be bounded strictly positive functions such that $\epsilon(\zeta) < k_0(\zeta)$ for $\zeta \in \mathcal{I}$. 
We henceforth drop the $\zeta$ dependence of these functions and write simply $\rho$, $\epsilon$, $k_0$ for $\rho(\zeta)$, $\epsilon(\zeta)$, $k_0(\zeta)$, respectively.

Let $\Gamma = \Gamma(\zeta)$ be a family of oriented contours parametrized by $\zeta \in \mathcal{I}$ and let $\hat{\Gamma} = \Gamma \cup \{k \, | \, |k \pm k_0| = \epsilon \}$ denote the union of $\Gamma$ with the circles of radius $\epsilon$ centered at $\pm k_0$ oriented counterclockwise. Assume that, for each $\zeta \in \mathcal{I}$:
\begin{itemize}

\item[($\Gamma$1)] $\Gamma$ and $\hat{\Gamma}$ are Carleson jump contours up to reorientation of a subcontour.

\item[($\Gamma$2)] $\Gamma$ contains the two crosses $\pm k_0 + X^{\epsilon}$ as oriented subcontours.

\item[($\Gamma$3)] $\Gamma$ is invariant as a set under the map $k \mapsto - \bar{k}$. Moreover, the orientation of $\Gamma$ is such that if $k$ traverses $\Gamma$ in the positive direction, then $-\bar{k}$ traverses $\Gamma$ in the negative direction.

\item[($\Gamma$4)] The point $\infty \in \hat{\C}$ can be approached nontangentially, i.e., there exists a sector $W_{a,b}$ which is a nontangential sector at $\infty$ for $\Gamma$.
\end{itemize}
Moreover, assume that the Cauchy singular operator $\mathcal{S}_{\hat{\Gamma}}$ defined by
\begin{align*}
(\mathcal{S}_{\hat{\Gamma}} h)(z) = \lim_{r \to 0} \frac{1}{\pi i} \int_{\hat{\Gamma} \setminus \{z' \, | \, |z' - z| < r\}} \frac{h(z')}{z' - z} dz',
\end{align*}
is uniformly\footnote{For any fixed $\zeta \in \mathcal{I}$, $\mathcal{S}_{\hat{\Gamma}}$ is bounded on $L^2(\hat{\Gamma})$ as a consequence of ($\Gamma$1), see e.g. \cite{LCarleson}.} bounded on $L^2(\hat{\Gamma})$, i.e. 
\begin{align}\label{cauchysingularbound}
\sup_{\zeta \in \mathcal{I}} \|\mathcal{S}_{\hat{\Gamma}}\|_{\mathcal{B}(L^2(\hat{\Gamma}))} < \infty.
\end{align}

Consider the following family of $L^2$-RH problems parametrized by the two parameters $\zeta \in \mathcal{I}$ and $t > 0$:
\begin{align}\label{RHm}
\begin{cases} m(\zeta, t, \cdot) \in I + \dot{E}^2(\hat{\C} \setminus \Gamma), \\
m_+(\zeta, t, k) = m_-(\zeta, t, k) v(\zeta, t, k) \quad \text{for a.e.} \ k \in \Gamma, 
\end{cases} 
\end{align}
where the jump matrix $v(\zeta, t, k)$ satisfies
\begin{align}\label{winL1L2Linf}
&w(\zeta, t,\cdot) := v(\zeta, t,\cdot) - I \in L^1(\Gamma) \cap L^2(\Gamma) \cap L^\infty(\Gamma), \qquad \zeta \in \mathcal{I}, \  t > 0,
	\\
&\det v(\zeta, t, \cdot) = 1 \;\; \text{a.e. on $\Gamma$}, \qquad \zeta \in \mathcal{I}, \  t > 0,  
\end{align}
and
\begin{align}\label{vsymm}
  v(\zeta,t,k) = \overline{v(\zeta, t, -\bar{k})}, \qquad \zeta \in \mathcal{I}, \  t > 0, \  k \in \Gamma.
\end{align}

Let $\tau := t \rho^2$. Let $\Gamma_X$ be the union of the two crosses $\pm k_0 + X^{\epsilon}$ and let $\Gamma' = \Gamma \setminus \Gamma_X$. Suppose 
\begin{align} \label{wL12infty}
\begin{cases}
 \|w(\zeta,t,\cdot)\|_{L^1(\Gamma')} = O(\epsilon \tau^{-1}), 
	\\ 
 \|w(\zeta,t,\cdot)\|_{L^\infty(\Gamma')} = O(\tau^{-1}),
\end{cases} \qquad \tau \to \infty, \  \zeta \in \mathcal{I}, 
\end{align}
uniformly with respect to $\zeta \in \mathcal{I}$. Moreover, suppose that near $k_0$ the normalized jump matrix
\begin{align}\label{vjdef}
v_0(\zeta,t,z) =  v\biggl(\zeta, t, k_0 + \frac{\epsilon z}{\rho}\biggr), \qquad z \in X^{\rho},
\end{align}
has the form
\begin{align}\label{smallcrossjump}
v_0(\zeta, t, z) = \begin{cases}
\begin{pmatrix} 1 & 0	\\
  R_1(\zeta, t, z)z^{-2i\nu(\zeta)} e^{t\phi(\zeta, z)}	& 1 \end{pmatrix}, &  z \in X_1^{\rho}, \\
\begin{pmatrix} 1 & -R_2(\zeta, t, z)z^{2i\nu(\zeta)}e^{-t\phi(\zeta, z)}	\\
0 & 1  \end{pmatrix}, &  z \in X_2^{\rho}, \\
\begin{pmatrix} 1 &0 \\
 -R_3(\zeta, t, z)z^{-2i\nu(\zeta)} e^{t\phi(\zeta, z)}	& 1 \end{pmatrix}, &  z \in X_3^{\rho},
 	\\
 \begin{pmatrix} 1	& R_4(\zeta, t, z)z^{2i\nu(\zeta)}e^{-t\phi(\zeta, z)}	\\
0	& 1 \end{pmatrix}, & z \in X_4^{\rho}, 
\end{cases}
\end{align}
where:
\begin{itemize}
\item The phase $\phi(\zeta, z)$ is a smooth function of $(\zeta, z) \in \mathcal{I} \times \C$ 
such that 
\begin{align}\label{phiassumptions}
\phi(\zeta, 0) \in i\R, \qquad \frac{\partial \phi}{\partial z}(\zeta, 0) = 0, \qquad \frac{\partial^2 \phi}{\partial z^2}(\zeta, 0) = i
\end{align}
for $\zeta \in \mathcal{I}$, and
\begin{subequations}
\begin{align} \label{rephiestimatea}
 & \re \phi(\zeta,z) \leq -\frac{|z|^2}{4}, &&  z \in X_1^{\rho} \cup X_3^{\rho}, \  \zeta \in \mathcal{I},
  	\\ \label{rephiestimateb}
 & \re \phi(\zeta,z) \geq \frac{|z|^2}{4}, &&  z \in X_2^{\rho} \cup X_4^{\rho},  \  \zeta \in \mathcal{I},
  	\\ \label{Phiz3estimate}
 & \biggl|\phi(\zeta, z) - \phi(\zeta,0) - \frac{iz^2}{2}\biggr| \leq C \frac{|z|^3}{\rho}, && z \in X^{\rho}, \  \zeta \in \mathcal{I},
\end{align}
\end{subequations}
where $C > 0$ is a constant.

\item There exist a function $q:\mathcal{I} \to \C$ with $\sup_{\zeta \in \mathcal{I}} |q(\zeta)| < 1$, and constants $(\alpha, L) \in [\frac{1}{2},1) \times (0,\infty)$ such that the functions $\{R_j(\zeta, t, z)\}_1^4$ satisfy the following inequalities:
\begin{align} \label{Lipschitzconditions}
\begin{cases}
   |R_1(\zeta, t, z) - q(\zeta)| \leq  L  \bigl| \frac{z}{\rho}\bigr|^\alpha e^{\frac{t|z|^2}{6}}, & z \in X_1^{\rho},  \\
 \left|R_2(\zeta, t, z) - \frac{\overline{q(\zeta)}}{1 - |q(\zeta)|^2}\right| \leq  L \bigl| \frac{z}{\rho}\bigr|^\alpha e^{\frac{t|z|^2}{6}}, \qquad& z \in X_2^{\rho}, \vspace{.1cm}\\ 
  \left|R_3(\zeta, t, z) - \frac{q(\zeta)}{1 - |q(\zeta)|^2}\right| \leq  L \bigl| \frac{z}{\rho}\bigr|^\alpha e^{\frac{t|z|^2}{6}}, & z \in X_3^{\rho}, 
  	\\
  |R_4(\zeta, t, z) - \overline{q(\zeta)}| \leq L  \bigl| \frac{z}{\rho}\bigr|^\alpha e^{\frac{t|z|^2}{6}}, & z \in X_4^{\rho}, 
\end{cases} \  \zeta \in \mathcal{I}, \  t > 0.
\end{align}
\item The function $\nu(\zeta)$ is defined by $\nu(\zeta) = -\frac{1}{2\pi} \ln(1 -  |q(\zeta)|^2)$.
\end{itemize}

Then the $L^2$-RH problem (\ref{RHm}) has a unique solution for all sufficiently large $\tau$ and this solution satisfies
\begin{align}\label{limlm12}
\ntlim_{k\to \infty} (k m(\zeta,t,k))_{12}
= -\frac{2i\epsilon \re{\beta(\zeta, t)}}{\sqrt{\tau}}  + O\bigl(\epsilon \tau^{-\frac{1+\alpha}{2}} \bigr), \qquad \tau \to \infty, \  \zeta \in \mathcal{I},
\end{align}
where the error term is uniform with respect to $\zeta \in \mathcal{I}$ and $\beta(\zeta, t)$ is defined by
\begin{align}\label{betadef}
  \beta(\zeta, t) = \sqrt{\nu(\zeta)} e^{i\left(\frac{\pi}{4} - \arg q(\zeta) + \arg \Gamma(i\nu(\zeta))\right)} e^{-t\phi(\zeta, 0)} t^{-i\nu(\zeta)}. 
\end{align}  
\end{theorem}

\begin{remark}\upshape\label{intuitionremark}
In order to provide some intuition for the statement of Theorem \ref{steepestdescentth}, we note that when applying this theorem to the mKdV equation in Section \ref{similaritysec}, we will identify $\zeta$ with the quotient $\zeta := x/t$ and choose $\mathcal{I} = [-N,0)$ with  $N > 0$.
The two critical points will be located at $\pm k_0$ with $k_0 := \sqrt{|\zeta|/12}$ and we will take $\epsilon := k_0/2$ and $\rho := \epsilon \sqrt{48 k_0}$ (see equation (\ref{choices})).
The contours $\Gamma$ and $\hat{\Gamma}$ will be as displayed in Figure \ref{Gammafig}.
Also, we will have $\tau = t\rho^2 =  \sqrt{|x|^3/(12 t)}$ and the phase function $\phi(\zeta, z)$ will be given by
$$\phi(\zeta, z) = -16i k_0^3 + \frac{iz^2}{2} + \frac{i z^3}{12 \rho}.$$
The idea is that the disks of radius $\epsilon$ centered at $\pm k_0$ provide small neighborhoods of the two critical points in the $k$-plane, such that as $t \to \infty$ the jump matrix is close to the identity matrix everywhere except on the two small crosses $ \pm k_0 + X^{\epsilon}$ (cf. assumption (\ref{wL12infty})). 
Near the critical point $k_0$, it is convenient to work with the variable $z = \frac{\rho(k -k_0)}{\epsilon}$ (i.e. $k = k_0 + \frac{\epsilon z}{\rho}$) which centers the critical point at the origin and maps the cross $k_0 + X^{\epsilon}$ of radius $\epsilon$ in the $k$-plane to the cross $X^\rho$ of radius $\rho$ in the $z$-plane. The scaling factor $\rho$ is chosen so that the jump matrix takes the form (\ref{smallcrossjump}) in the $z$-plane. This form of the jump matrix together with the estimates in (\ref{Lipschitzconditions}) make it possible to relate the solution near the critical point $k_0$ to the solution of the model RH problem studied in Appendix \ref{exactapp}.
The variable $\tau$ is introduced so that $\tau \to \infty$ corresponds to the condition $|x|^3/t \to \infty$ relevant for the similarity sector (cf. equation (\ref{similaritysector})).
\end{remark}

\begin{remark}\upshape
The conclusion of Theorem \ref{steepestdescentth} can be stated more explicitly as follows: There exist constants $T> 0$ and $K>0$ such that the solution $m(\zeta, t,k)$ of (\ref{RHm}) exists, the nontangential limit $\ntlim_{k\to \infty}(k m(\zeta,t,k))_{12}$ exists, and the inequality
\begin{align}\nonumber
  \left| \lim_{k\to \infty} (k m(\zeta,t,k))_{12} + \frac{2i\epsilon\re{\beta(\zeta, t)}}{\sqrt{\tau}}\right|
  \leq K \epsilon \tau^{-\frac{1+\alpha}{2}}
\end{align}
holds for all $(\zeta, t) \in \mathcal{I} \times [0, \infty)$ such that $\tau = t \rho^2 > T$.
\end{remark}

\begin{remark}\upshape
Theorem \ref{steepestdescentth} allows for jump matrices $v(\zeta,t,k)$ that depend discontinuously on $(\zeta,t)$. It also allows for contours $\Gamma$ and functions $\rho, \epsilon, k_0$ that depend discontinuously on $\zeta$.
\end{remark}

\subsection{Proof of Theorem \ref{steepestdescentth}}
Since $\det v = 1$, uniqueness of $m$ follows from Lemma \ref{uniquelemma}.

Let $m^X$ be the solution of Theorem \ref{crossth} and let
\begin{align*}
& D(\zeta, t) = \begin{pmatrix} e^{-\frac{t\phi(\zeta, 0)}{2}}t^{-\frac{i\nu(\zeta)}{2}} & 0 \\ 
0 & e^{\frac{t\phi(\zeta, 0)}{2}}t^{\frac{i\nu(\zeta)}{2}}  \end{pmatrix}.
\end{align*}
Define $m_0(\zeta,t,k)$ near $k = k_0$ by
\begin{align*}
& m_0(\zeta, t, k) =
D(\zeta, t) m^X\Bigl(q(\zeta), \frac{\sqrt{\tau}}{\epsilon} (k-k_0)\Bigr) D(\zeta, t)^{-1}, \qquad |k - k_0| \leq \epsilon, 
\end{align*}
and extend it to a neighborhood of $k = -k_0$ by symmetry:
\begin{align}\label{m0symm}
m_0(\zeta,t,k) = \overline{m_0(\zeta, t, -\bar{k})}, \qquad |k + k_0| \leq \epsilon.
\end{align}
Lemma \ref{deformationlemma} implies that $m$ satisfies the $L^2$-RH problem (\ref{RHm}) if and only if the function $\hat{m}(\zeta, t, k)$ defined by
\begin{align*}
\hat{m}(\zeta,t,k) = \begin{cases}
m(\zeta, t, k)m_0(\zeta,t,k)^{-1}, & |k \pm k_0| < \epsilon,\\
m(\zeta, t, k), & \text{otherwise},
\end{cases}
\end{align*}
satisfies the $L^2$-RH problem
\begin{align}\label{RHmhat}
\begin{cases}
\hat{m}(\zeta,t,\cdot) \in I + \dot{E}^2(\hat{\C} \setminus \hat{\Gamma}), \\
\hat{m}_+(\zeta,t,k) = \hat{m}_-(\zeta, t, k) \hat{v}(x, t, k) \quad \text{for a.e.} \ k \in \hat{\Gamma}, 
\end{cases}
\end{align}
where the jump matrix $\hat{v}$ is given by
\begin{align}\label{hatvdef}
\hat{v}(\zeta, t, k) 
=  \begin{cases}
 m_{0-}(\zeta, t, k) v(\zeta, t, k) m_{0+}(\zeta,t,k)^{-1}, & k \in \Gamma \cap \{|k \pm k_0| < \epsilon\}, \\
m_0(\zeta, t, k)^{-1}, & |k \pm k_0| = \epsilon, \\
v(\zeta, t, k),  & \text{otherwise}.
\end{cases}
\end{align}
It follows from (\ref{vsymm}) and (\ref{m0symm}) that $\hat{w} = \hat{v} - I$ obeys the symmetry
\begin{align} \label{whatsymm}
   \hat{w}(\zeta, t, k) = \overline{\hat{w}(\zeta,t,-\bar{k})}, \qquad k \in \hat{\Gamma}.
\end{align}

\begin{lemma}\label{lemma1}
The function $\hat{w} = \hat{v} - I$ satisfies
\begin{align}\label{hatwestimate}
\hat{w}(\zeta,t,k) = O\bigl(\tau^{-\frac{\alpha}{2}} e^{-\frac{\tau}{24\epsilon^2}|k \mp k_0|^2}\bigr), \qquad \tau \to \infty, \  \zeta \in \mathcal{I}, \  k \in \pm k_0 + X^\epsilon,
\end{align}
where the error term is uniform with respect to $(\zeta, k)$ in the given ranges.
\end{lemma}
\begin{proof}
We assume $k \in k_0 + X^\epsilon$; the case of $k \in -k_0 + X^\epsilon$ follows by symmetry. Then
\begin{align*}
\hat{w}(\zeta, t,k)
& = m_{0-}(\zeta, t, k) v(\zeta, t, k) m_{0+}(\zeta,t,k)^{-1} - I
	\\
& = m_{0-}(\zeta, t, k)  u(\zeta,t,k) m_{0+}(\zeta,t,k)^{-1},
\end{align*}
where
$$
u(\zeta,t,k) := v(\zeta, t, k) -  D(\zeta,t) v^X\biggl(q(\zeta), \frac{\sqrt{\tau}}{\epsilon}(k-k_0)\biggr) D(\zeta,t)^{-1}$$
and the jump matrix $v^X$ is defined in Appendix \ref{exactapp}.
The functions $m_{0+}(\zeta, t, k)$ and $m_{0-}(\zeta, t, k)$ are uniformly bounded for $t>0$, $\zeta \in \mathcal{I}$, and $k \in k_0 + X^\epsilon$ by equation (\ref{mXbounded}). Therefore, it is enough to prove that 
\begin{align}\label{Ujbound}
u(\zeta,t,k) = O\bigl(\tau^{-\frac{\alpha}{2}} e^{-\frac{\tau}{24\epsilon^2}|k-k_0|^2}\bigr), \qquad \tau \to \infty, \  \zeta \in \mathcal{I}, \  k \in k_0 + X^\epsilon,
\end{align}
uniformly with respect to $(\zeta, k)$.
Introducing the function $u_0$ by
$$u_0(\zeta,t,z) =  u\biggl(\zeta,t,k_0 + \frac{\epsilon z}{\rho}\biggr) 
= v_0(\zeta, t, z) - D(\zeta,t) v^X(q(\zeta), \sqrt{t} z) D(\zeta,t)^{-1},$$
we can rewrite the condition (\ref{Ujbound}) as follows:
\begin{align}\label{ujbound}
u_0(\zeta,t,z) = O\bigl(\tau^{-\frac{\alpha}{2}} e^{-\frac{t|z|^2}{24}}\bigr), \qquad \tau \to \infty, \  \zeta \in \mathcal{I}, \  z \in X^{\rho},
\end{align}
uniformly with respect to $(\zeta, z)$ in the given ranges.
Using that
\begin{align*}
D(\zeta,t) v^X\bigl(q(\zeta), \sqrt{t} z\bigr) D(\zeta,t)^{-1} = \begin{cases}
 \begin{pmatrix} 1 & 0 \\
  q(\zeta)  z^{-2i\nu(\zeta)} e^{\frac{itz^2}{2}}e^{ t\phi(\zeta, 0)}	& 1 \end{pmatrix}, &   z \in X_1, 
  	\\
\begin{pmatrix} 1 & - \frac{\overline{q(\zeta)}}{1 -  |q(\zeta)|^2} z^{2i\nu(\zeta)}e^{-\frac{itz^2}{2}}e^{-t\phi(\zeta, 0)} 	\\
0 & 1  \end{pmatrix}, &  z \in X_2, 
	\\
\begin{pmatrix} 1 &0 \\
- \frac{q(\zeta)}{1 - |q(\zeta)|^2} z^{-2i\nu(\zeta)} e^{\frac{itz^2}{2}}e^{t\phi(\zeta, 0)} & 1 \end{pmatrix}, &  z \in X_3,
	\\
\begin{pmatrix} 1 &  \overline{q(\zeta)} z^{2i\nu(\zeta)}e^{-\frac{itz^2}{2}} e^{-t\phi(\zeta, 0)} \\
0 & 1  \end{pmatrix}, &  z \in X_4,	
\end{cases}
\end{align*}
equation (\ref{ujbound}) follows from the assumptions (\ref{smallcrossjump})-(\ref{Lipschitzconditions}). Indeed, we will give the details of the proof of (\ref{ujbound}) in the case of $z \in X_1^{\rho}$; the other cases are similar.

Let $z \in X_1^{\rho}$. In this case only the $(21)$ entry of $u_0(\zeta,t,z)$ is nonzero and using that $\arg z = \frac{\pi}{4}$ and $\sup_{\zeta \in \mathcal{I}} |q(\zeta)| < 1$, we find
\begin{align} \nonumber
|(u_0(\zeta, t, z))_{21}| = &\;
\big|R_1(\zeta, t, z)z^{-2i\nu(\zeta)}e^{t\phi(\zeta, z)}
- q(\zeta) z^{-2i\nu(\zeta)}e^{\frac{itz^2}{2}} e^{t\phi(\zeta, 0)}\big|
	\\ \nonumber
= & \; |z^{-2i\nu(\zeta)}| \bigl|R_1(\zeta, t, z)e^{t\hat{\phi}(\zeta, z)} - q(\zeta)\bigr| |e^{t\phi(\zeta, 0)}| e^{-\frac{t|z|^2}{2}} 
  	\\ \nonumber
 \leq &\; e^{\frac{\pi \nu(\zeta)}{2}}\Bigl(\bigl|R_1(\zeta, t, z) - q(\zeta)\bigr|e^{t\re \hat{\phi}(\zeta, z)}
 + |q(\zeta)| \bigl|e^{t\hat{\phi}(\zeta, z)} - 1\bigr|\Bigr)
 e^{-\frac{t|z|^2}{2}}, 
 	\\ \label{Westimate}
& \hspace{5cm} \zeta \in \mathcal{I}, \  t > 0, \  z \in X_1^{\rho},
\end{align}
where $\hat{\phi}(\zeta, z) = \phi(\zeta,z)- \phi(\zeta, 0) - \frac{iz^2}{2}$.
The simple estimate
$$|e^w -1| = \biggl|\int_0^1 w e^{sw} ds\biggr| \leq |w| \max_{s \in [0,1]}e^{s\re w}, \qquad w \in \C,$$
yields the inequality
\begin{align}\label{ewminus1estimate}  
  |e^w - 1| \leq |w| \max(1, e^{\re w}), \qquad w \in \C.
\end{align}
On the other hand, by (\ref{phiassumptions}) and (\ref{rephiestimatea}),
\begin{align}\label{rehatphi}
\re \hat{\phi}(\zeta, z) = \re \phi(\zeta,z) + \frac{|z|^2}{2} \leq \frac{|z|^2}{4}, \qquad \zeta \in \mathcal{I}, \  z \in X_1^{\rho}.
\end{align}
Using (\ref{ewminus1estimate}), (\ref{rehatphi}), and the fact that $\sup_{\zeta \in \mathcal{I}} |q(\zeta)| < 1$ in (\ref{Westimate}), we find
\begin{align*}
 |(u_0(\zeta, t, z))_{21}| \leq &\;  e^{\frac{\pi \nu(\zeta)}{2}}\Bigl(\bigl|R_1(\zeta, t, z) - q(\zeta)\bigr| 
  + t\bigl|\hat{\phi}(\zeta, z)\bigr|\Bigr) e^{-\frac{t|z|^2}{4}}, 
  	\\
& \hspace{5cm}   \zeta \in \mathcal{I}, \  t > 0,\  z \in X_1^{\rho}. 
\end{align*}  
By (\ref{Phiz3estimate}), (\ref{Lipschitzconditions}), and the fact that $\sup_{\zeta \in \mathcal{I}} |\nu(\zeta)| < \infty$, the right-hand side is of order
\begin{align} \nonumber
& O\biggl(\biggl( \frac{L |z|^\alpha e^{\frac{t|z|^2}{6}}}{\rho^\alpha} + \frac{tC|z|^3}{\rho} \biggr) e^{-\frac{t|z|^2}{4}}  \biggr)
= O\biggl( \biggl(\frac{(t|z|^2)^{\alpha/2}}{\tau^{\alpha/2}} + \frac{(t|z|^2)^{3/2}}{\tau^{1/2}}\biggr) e^{-\frac{t|z|^2}{12}}\biggr) 
 	\\
&  = O\biggl( \biggl(\frac{1}{\tau^{\alpha/2}} + \frac{1}{\tau^{1/2}}\biggr) e^{-\frac{t|z|^2}{24}}\biggr), \qquad \tau \to \infty, \  \zeta \in \mathcal{I}, \  z \in X_1^{\rho},
\end{align}
uniformly with respect to $(\zeta, z)$ in the given ranges. This proves (\ref{ujbound}) in the case of $z \in X_1^{\rho}$.
\end{proof}

\begin{lemma}\label{lemma2}
 We have
\begin{subequations}
\begin{align} \label{hatwestimatea}
&\|\hat{w}(\zeta, t, \cdot)\|_{L^2(\hat{\Gamma})} = O(\epsilon^{\frac{1}{2}} \tau^{-\frac{\alpha}{2}}), \qquad \tau \to \infty, 
\  \zeta \in \mathcal{I}, 
	\\ \label{hatwestimateb}
&\|\hat{w}(\zeta, t, \cdot)\|_{L^\infty(\hat{\Gamma})} = O(\tau^{-\frac{\alpha}{2}}), \qquad \tau \to \infty, 
\  \zeta \in \mathcal{I},
\end{align}
\end{subequations}
and, for any $p \in [1, \infty)$,
\begin{align}\label{hatwestimatec}
& \|\hat{w}(\zeta, t, \cdot)\|_{L^p(\pm k_0 + X^\epsilon)} = O(\epsilon^{\frac{1}{p}} \tau^{-\frac{1}{2p} - \frac{\alpha}{2}}), \qquad \tau \to \infty, \  \zeta \in \mathcal{I},
\end{align}
where all error terms are uniform with respect to $\zeta$.
\end{lemma}
\begin{proof}
We have
\begin{align}\nonumber
\|\hat{w}(\zeta, t, \cdot)\|_{L^2(\hat{\Gamma})} 
\leq   & \; \|\hat{w}(\zeta, t, \cdot)\|_{L^2(\Gamma')} + \|m_0(\zeta, t, \cdot)^{-1} - I \|_{L^2(|k - k_0| = \epsilon)} 
	\\ \nonumber
&+ \|m_0(\zeta, t, \cdot)^{-1} - I \|_{L^2(|k + k_0| = \epsilon)}  + \|\hat{w}(\zeta, t, \cdot) \|_{L^2(k_0 + X^\epsilon)} 
	\\ \label{whatnorm}
& + \|\hat{w}(\zeta, t, \cdot) \|_{L^2(-k_0 + X^\epsilon)}.
\end{align}

On $\Gamma'$, the matrix $\hat{w}$ is given by either $w$ or $m_0wm_0^{-1}$. Since the assumption (\ref{wL12infty}) implies 
$$\|w(\zeta,t,\cdot)\|_{L^2(\Gamma')} \leq \sqrt{\|w(\zeta,t,\cdot)\|_{L^\infty(\Gamma')} \|w(\zeta,t,\cdot)\|_{L^1(\Gamma')}}
= O(\epsilon^{\frac{1}{2}} \tau^{-1}), \qquad \tau \to \infty, \  \zeta \in \mathcal{I},$$
this gives $\|\hat{w}(\zeta, t, \cdot)\|_{L^2(\Gamma')} = O(\epsilon^{\frac{1}{2}} \tau^{-1})$.
Moreover, by (\ref{mcasymptotics}), $m^X(q, z) = I + O\bigl(\frac{1}{z}\bigr)$ as $z \to \infty$ uniformly with respect to $\arg z \in [0, 2\pi]$ and $q$ in compact subsets of $\D$. Hence, as the entries of $D(\zeta, t)$ have unit modulus,
\begin{align}\nonumber
&\|m_0(\zeta, t, k)^{-1} - I \|_{L^p(|k - k_0| = \epsilon)} 
 	\\ \nonumber
& = 
\biggl \| D(\zeta, t) \biggl[m^X\biggl(q(\zeta),  \frac{\sqrt{\tau}}{\epsilon}( k - k_0)\biggr)^{-1}  - I \biggr]D(\zeta, t)^{-1} \biggr \|_{L^p(|k - k_0| = \epsilon)}  
	\\ \label{star}
& = \begin{cases} O(\epsilon^{1/p} \tau^{-1/2}), & p \in [1,\infty), \\ 
O(\tau^{-1/2}), & p = \infty,
\end{cases}
\end{align}
uniformly with respect to $\zeta \in \mathcal{I}$; the third term on the right-hand side of (\ref{whatnorm}) satisfies a similar estimate. The last two terms in (\ref{whatnorm}) are easily estimated using (\ref{hatwestimate}). This yields (\ref{hatwestimatea}). The proof of (\ref{hatwestimateb}) is similar.

In order to prove (\ref{hatwestimatec}), we note that (\ref{hatwestimate}) implies
\begin{align} \nonumber
  \|\hat{w}(\zeta, t, \cdot)\|_{L^p(k_0 + X^\epsilon)} 
 & = O\biggl(  \tau^{-\frac{\alpha}{2}} \bigg(\int_{k_0 + X^\epsilon} e^{-\frac{p\tau}{24\epsilon^2} |k-k_0|^2} |dk|\bigg)^{\frac{1}{p}} \biggr)
  	\\ \label{hatwL1estimate}
& = O\biggl(\tau^{-\frac{\alpha}{2}}  \bigg(\int_0^\epsilon e^{-\frac{p\tau}{24\epsilon^2} u^2} du\bigg)^{\frac{1}{p}} \biggr), \qquad \tau \to \infty, \  \zeta \in \mathcal{I}.
\end{align}
Letting $v = \frac{p\tau}{24\epsilon^2} u^2$ we find 
\begin{align}\label{int0epsilonestimate}
\int_0^\epsilon e^{-\frac{p\tau}{24\epsilon^2} u^2} du
\leq \int_0^\infty  e^{-\frac{p\tau}{24\epsilon^2} u^2} du 
= \frac{\epsilon\sqrt{6}}{\sqrt{p\tau}} \int_0^\infty \frac{e^{-v}}{\sqrt{v}} dv = \frac{\epsilon \sqrt{6\pi}}{\sqrt{p\tau}}.
\end{align}
Equations (\ref{hatwL1estimate}) and (\ref{int0epsilonestimate}) yield (\ref{hatwestimatec}).
\end{proof}

Let $\hat{\mathcal{C}}$ denote the Cauchy operator associated with $\hat{\Gamma}$:
$$(\hat{\mathcal{C}} f)(z) = \frac{1}{2\pi i} \int_{\hat{\Gamma}} \frac{f(s)}{s - z} ds, \qquad z \in \C \setminus \hat{\Gamma}.$$
We define $\hat{\mathcal{C}}_{\hat{w}}: L^2(\hat{\Gamma}) + L^\infty(\hat{\Gamma}) \to L^2(\hat{\Gamma})$ by
$\hat{\mathcal{C}}_{\hat{w}}f = \hat{\mathcal{C}}_-(f \hat{w})$, i.e. $\hat{\mathcal{C}}_{\hat{w}}$ is defined by equation (\ref{Cwdef}) where we have chosen, for simplicity, $\hat{w}^+ = \hat{w}$ and $\hat{w}^- = 0$.

\begin{lemma}\label{lemma3}
There exists a $T > 0$ such that $I - \hat{\mathcal{C}}_{\hat{w}(\zeta, t, \cdot)} \in \mathcal{B}(L^2(\hat{\Gamma}))$ is invertible for all $(\zeta,t) \in \mathcal{I} \times (0, \infty)$ with $\tau > T$.
\end{lemma}
\begin{proof}
Assumption (\ref{cauchysingularbound}) together with the Sokhotski-Plemelj formula $\hat{\mathcal{C}}_- = \frac{1}{2}(-I + \mathcal{S}_{\hat{\Gamma}})$ show that $\sup_{\zeta \in \mathcal{I}} \|\hat{\mathcal{C}}_-\|_{\mathcal{B}(L^2(\hat{\Gamma}))} < \infty$.
Thus, by (\ref{Cwnorm}) and (\ref{hatwestimateb}),
\begin{align}\label{Chatwnorm}
\|\hat{\mathcal{C}}_{\hat{w}}\|_{\mathcal{B}(L^2(\hat{\Gamma}))} \leq C \|\hat{w}\|_{L^\infty(\hat{\Gamma})}  
= O(\tau^{-\frac{\alpha}{2}}), \qquad \tau \to \infty, \  \zeta \in \mathcal{I},
\end{align}
uniformly with respect to $\zeta$. This proves the lemma. 
\end{proof}

In view of Lemma \ref{lemma3}, we may define the $2 \times 2$-matrix valued function $\hat{\mu}(\zeta, t, z)$ whenever $\tau > T$ by
\begin{align}\label{hatmudef}
\hat{\mu} = I + (I - \hat{\mathcal{C}}_{\hat{w}})^{-1}\hat{\mathcal{C}}_{\hat{w}}I  \in I + L^2(\hat{\Gamma}).
\end{align}

\begin{lemma}\label{lemma4}
The function $\hat{\mu}(\zeta, t, k)$ satisfies
\begin{align}\label{muhatestimate}
\|\hat{\mu}(\zeta,t,\cdot) - I\|_{L^2(\hat{\Gamma})} = O\big(\epsilon^{\frac{1}{2}}\tau^{-\frac{\alpha}{2}}\big), \qquad \tau \to \infty, \  \zeta \in \mathcal{I},
\end{align}
where the error term is uniform with respect to $\zeta \in \mathcal{I}$.
\end{lemma}
\begin{proof}
Utilizing the Neumann series representation
\begin{align}\label{neumannseries}
(I - \hat{\mathcal{C}}_{\hat{w}})^{-1} = \sum_{j=0}^\infty \hat{\mathcal{C}}_{\hat{w}}^j
\end{align}
we obtain
\begin{align*}
\|(I - \hat{\mathcal{C}}_{\hat{w}})^{-1}\|_{\mathcal{B}(L^2(\hat{\Gamma}))} 
& \leq \sum_{j=0}^\infty \|\hat{\mathcal{C}}_{\hat{w}}\|_{\mathcal{B}(L^2(\hat{\Gamma}))}^j
 = \frac{1}{1 - \|\hat{\mathcal{C}}_{\hat{w}}\|_{\mathcal{B}(L^2(\hat{\Gamma}))}}
\end{align*}
whenever $\|\hat{\mathcal{C}}_{\hat{w}}\|_{\mathcal{B}(L^2(\hat{\Gamma}))} < 1$.
Using the bound $\sup_{\zeta \in \mathcal{I}} \|\hat{\mathcal{C}}_-\|_{\mathcal{B}(L^2(\hat{\Gamma}))} < \infty$ and equation (\ref{Chatwnorm}), we find
\begin{align*}
\|\hat{\mu} - I\|_{L^2(\hat{\Gamma})} & = 
\|(I- \hat{\mathcal{C}}_{\hat{w}})^{-1}\hat{\mathcal{C}}_{\hat{w}}I\|_{L^2(\hat{\Gamma})} 
 \leq \|(I - \hat{\mathcal{C}}_{\hat{w}})^{-1}\|_{\mathcal{B}(L^2(\hat{\Gamma}))}
\|\hat{\mathcal{C}}_-(\hat{w})\|_{L^2(\hat{\Gamma})}
	\\
& \leq  
\frac{C\|\hat{w}\|_{L^2(\hat{\Gamma})}}{1 - \|\hat{\mathcal{C}}_{\hat{w}}\|_{\mathcal{B}(L^2(\hat{\Gamma}))}}
\leq C\|\hat{w}\|_{L^2(\hat{\Gamma})}
\end{align*}
for all $\zeta \in \mathcal{I}$ and all $\tau$ large enough. In view of (\ref{hatwestimatea}), this gives (\ref{muhatestimate}).
\end{proof}

\begin{lemma}\label{lemma5}
There exists a unique solution $\hat{m} \in I + \dot{E}^2(\hat{\C} \setminus \hat{\Gamma})$ of the $L^2$-RH problem (\ref{RHmhat}) whenever $\tau > T$. This solution is given by
\begin{align}\label{hatmrepresentation}
\hat{m}(\zeta, t, k) = I + \hat{\mathcal{C}}(\hat{\mu}\hat{w}) = I + \frac{1}{2\pi i}\int_{\hat{\Gamma}} \hat{\mu}(\zeta, t, s) \hat{w}(\zeta, t, s) \frac{ds}{s - k}.
\end{align}
\end{lemma}
\begin{proof}
Uniqueness follows from Lemma \ref{uniquelemma} since $\det \hat{v} = 1$. Moreover, equation (\ref{hatmudef}) implies that $\hat{\mu} - I = \hat{\mathcal{C}}_{\hat{w}} \hat{\mu}$. Hence, by Lemma \ref{mulemma}, $\hat{m} = I + \hat{\mathcal{C}}(\hat{\mu} \hat{w})$ satisfies the $L^2$-RH problem (\ref{RHmhat}). 
\end{proof}

\begin{lemma}\label{lemma6}
For each point $(\zeta, t) \in \mathcal{I} \times (0, \infty)$ with $\tau > T$, the nontangential limit of $k(\hat{m}(\zeta,t,k) - I)$ as $k \to \infty$ exists and is given by
\begin{align}\label{limlhatm}
\ntlim_{k\to \infty} k(\hat{m}(\zeta,t,k) - I) 
=  - \frac{1}{2\pi i}\int_{\hat{\Gamma}} \hat{\mu}(\zeta,t,k) \hat{w}(\zeta,t,k) dk.
\end{align}
\end{lemma}
\begin{proof}
If $k \to \infty$ within a nontangential sector $W_{a,b} = \{a \leq \arg k \leq b\}$, then a simple argument shows that there exists a $c>0$ such that $|s-k| > c(|s| + |k|)$ for all $s \in \hat{\Gamma}$ and all $k \in W_{a,b}$ with $|k|$ sufficiently large.
Thus, since $\hat{\mu}\hat{w} \in L^1(\hat{\Gamma})$, dominated convergence implies
\begin{align*}
& \ntlim_{k\to \infty} \bigg|\int_{\hat{\Gamma}} (\hat{\mu} \hat{w})(\zeta, t, s) \frac{k ds}{s - k} + \int_{\hat{\Gamma}} (\hat{\mu} \hat{w})(\zeta, t, s) ds\bigg| 
\leq \ntlim_{k\to \infty} \int_{\hat{\Gamma}} |(\hat{\mu} \hat{w})(\zeta, t, s)| \frac{|s| |ds|}{|s - k|} 
	\\
& 
\leq \ntlim_{k\to \infty} \int_{\hat{\Gamma}} |(\hat{\mu} \hat{w})(\zeta, t, s)| \frac{|s| |ds|}{c(|s| + |k|)} 
= 0.
\end{align*}
The lemma now follows from (\ref{hatmrepresentation}).
\end{proof}

Equation (\ref{limlhatm}) implies that
\begin{align}\nonumber
& \ntlim_{k\to \infty} k(m(\zeta,t,k) - I) = \ntlim_{k\to \infty} k(\hat{m}(\zeta,t,k) - I) 
	\\ \nonumber
& = - \frac{1}{2\pi i}\biggl( \int_{|k -k_0| = \epsilon} + \int_{|k+k_0| = \epsilon}\biggr) \hat{\mu}(\zeta,t,k) \hat{w}(\zeta,t,k) dk
 - \frac{1}{2\pi i}\int_{\Gamma} \hat{\mu}(\zeta,t,k) \hat{w}(\zeta,t,k) dk.
\end{align}
By ($\Gamma$3) and (\ref{whatsymm}), if $f$ obeys the symmetry $f(k) = \overline{f(-\bar{k})}$, then so does $\hat{\mathcal{C}}_{\hat{w}}(f)$. In view of (\ref{neumannseries}), this implies that the operator $(I - \hat{\mathcal{C}}_{\hat{\mu}})^{-1}$ also preserves this symmetry. Thus $\hat{\mu}(\zeta, t, k) = \overline{\hat{\mu}(\zeta, t, -\bar{k})}$.
Together with (\ref{whatsymm}) this yields
$$ \int_{|k+k_0| = \epsilon} \hat{\mu}(\zeta,t,k) \hat{w}(\zeta,t,k)dk
= \overline{ \int_{|k-k_0| = \epsilon} \hat{\mu}(\zeta,t,k) \hat{w}(\zeta,t,k)dk }.$$
Hence, recalling that $\hat{w} = m_0^{-1} - I$ on the circle $|k -k_0| = \epsilon$,
\begin{align} \nonumber
\ntlim_{k\to \infty} k(m(\zeta,t,k) - I) 
= & - \frac{1}{\pi i}\re\biggl( \int_{|k - k_0| = \epsilon} \hat{\mu}(\zeta,t,k) (m_0(\zeta,t,k)^{-1} - I) dk\biggr)
	\\ \label{limlmminusI}
& - \frac{1}{2\pi i}\int_{\Gamma} \hat{\mu}(\zeta,t,k) \hat{w}(\zeta,t,k) dk.
\end{align}
By (\ref{mcasymptotics}),
\begin{align} \nonumber
  m_0(\zeta, t, k)^{-1} = &\; D(\zeta, t) m^X\biggl(q(\zeta), \frac{\sqrt{\tau}}{\epsilon}(k-k_0)\biggr)^{-1} D(\zeta, t)^{-1}
  	\\ \label{mjinvasymptotics}
= &\; I + \frac{B(\zeta, t)}{\sqrt{\tau}(k-k_0)} + O(\tau^{-1}), \qquad \tau \to \infty, \  \zeta \in \mathcal{I}, \  |k-k_0| = \epsilon,
\end{align}
where $B(\zeta, t)$ is defined by
\begin{align*}
B(\zeta, t) = -i\epsilon \begin{pmatrix} 0 & -\beta^X(q(\zeta))e^{-t\phi(\zeta,0)}t^{-i\nu(\zeta)} \\ \overline{\beta^X(q(\zeta))} e^{t\phi(\zeta,0)}t^{i\nu(\zeta)} & 0 \end{pmatrix}.
\end{align*}
Using (\ref{muhatestimate}) and (\ref{mjinvasymptotics}) we find
\begin{align}\nonumber
 \int_{|k -k_0| = \epsilon} & \hat{\mu}(\zeta,t,k) (m_0(\zeta,t,k)^{-1} - I) dk
= \int_{|k - k_0| = \epsilon} (m_0(\zeta,t,k)^{-1} - I) dk
	\\\nonumber
& + \int_{|k - k_0| = \epsilon}  (\hat{\mu}(\zeta,t,k) - I) (m_0(\zeta,t,k)^{-1} - I) dk
 	\\\nonumber
= &\; \frac{B(\zeta, t)}{\sqrt{\tau}} \int_{|k - k_0| = \epsilon} \frac{dk}{k - k_0} 
 + O\bigl(\epsilon \tau^{-1} \bigr) + O\biggl(\|\hat{\mu}(\zeta,t,\cdot) - I\|_{L^2(\hat{\Gamma})}\epsilon^{\frac{1}{2}}\tau^{-\frac{1}{2}} \biggr)
	\\ \label{inthatmumj}
= &\; \frac{2\pi iB(\zeta, t)}{\sqrt{\tau}} + O\bigl(\epsilon \tau^{-\frac{1+\alpha}{2}} \bigr), \qquad \tau \to \infty, \  \zeta \in \mathcal{I},
\end{align}
uniformly with respect to $\zeta \in \mathcal{I}$. 

On the other hand, 
\begin{align}\nonumber
\biggl|\int_{\Gamma} \hat{\mu}(\zeta,t,k) \hat{w}(\zeta,t,k) dk\biggr|
 & = \biggl|\int_{\Gamma}( \hat{\mu}(\zeta,t,k) -I) \hat{w}(\zeta,t,k) dk + \int_{\Gamma} \hat{w}(\zeta,t,k) dk\biggr|
	\\ \nonumber
& \leq \|\hat{\mu} - I\|_{L^2(\Gamma)}  \|\hat{w}\|_{L^2(\Gamma)} + \|\hat{w}\|_{L^1(\Gamma)}.
\end{align}
The $L^1$-norm of $\hat{w}$  is $O(\epsilon \tau^{-1})$ on $\Gamma'$ by (\ref{wL12infty}) and is $O(\epsilon\tau^{-\frac{1+\alpha}{2}})$ on $\{\pm k_0 + X^\epsilon\}$ by (\ref{hatwestimatec}). Hence $\|\hat{w}\|_{L^1(\Gamma)} = O(\epsilon \tau^{-\frac{1+\alpha}{2}})$.
Similarly, $\|\hat{w}\|_{L^2(\Gamma)} = O(\epsilon^{1/2} \tau^{-1} + \epsilon^{\frac{1}{2}} \tau^{-\frac{1}{4} -\frac{\alpha}{2}})$ by (\ref{wL12infty}) and (\ref{hatwestimatec}). Since $\|\hat{\mu} - I\|_{L^2(\Gamma)} = O(\epsilon^{1/2}\tau^{-\frac{\alpha}{2}})$ by (\ref{muhatestimate}) and $1/2 \leq \alpha < 1$, we infer that
\begin{align}\label{intSigmahatmuhatw}
\biggl|\int_{\Gamma}& \hat{\mu}(\zeta,t,k) \hat{w}(\zeta,t,k) dk\biggr| = O(\epsilon \tau^{-\frac{1+\alpha}{2}}), \qquad \tau \to \infty, \  \zeta \in \mathcal{I},
\end{align}
uniformly with respect to $\zeta \in \mathcal{I}$.
Equations (\ref{limlmminusI}), (\ref{inthatmumj}), and (\ref{intSigmahatmuhatw}) imply (\ref{limlm12}). This completes the proof of Theorem \ref{steepestdescentth}.

\section{Inverse scattering for the mKdV equation}\label{mkdvsec}\nequation
Before we apply Theorem \ref{steepestdescentth} to derive asymptotic formulas for the mKdV equation (\ref{mkdv}), we need to review how the solution of (\ref{mkdv}) with initial data $u_0(x)$ can be expressed in terms of the solution of a Riemann-Hilbert problem. 

Let
$$\mathsf{U}_0(x) = \begin{pmatrix} 0 & u_0(x) \\ u_0(x) & 0 \end{pmatrix}, \qquad \sigma_3 = \begin{pmatrix} 1 & 0 \\ 0 & -1 \end{pmatrix}.$$
Let $\mathsf{X}^+(x,k)$ and $\mathsf{X}^-(x,k)$ be the $2 \times 2$-matrix valued solutions of the linear Volterra integral equations
\begin{align}\label{mujdef}  
  \mathsf{X}^\pm(x,k) = I + \int_{\pm\infty}^x e^{i k(x'-x)\hat{\sigma}_3} (\mathsf{U}_0\mathsf{X}^\pm)(x,k) dx,
\end{align}
where $\hat{\sigma}_3$ acts on a $2\times 2$ matrix $A$ by $\hat{\sigma}_3A = [\sigma_3, A]$, i.e. $e^{\hat{\sigma}_3} A = e^{\sigma_3} A e^{-\sigma_3}$.
Define the spectral function $r(k)$ by 
\begin{align}\label{rdef}
  r(k) = -\frac{ \overline{b(\bar{k})}}{a(k)}, \qquad k \in \R,
\end{align}
where $a(k)$  and $b(k)$  are determined by the relation
\begin{align}\label{lineseq} 
  \mathsf{X}^+(x,k) = \mathsf{X}^-(x,k)e^{-ik x \hat{\sigma}_3} \begin{pmatrix} 
\overline{a(\bar{k})} 	&	b(k)	\\
\overline{b(\bar{k})}	&	a(k)
\end{pmatrix}, \qquad k \in \R.
\end{align}
The inverse scattering transform formalism \cite{AC1991, FT2007, BDT1988, IN1986} implies that the solution  $u(x,t)$  of (\ref{mkdv}) with initial data $u(x,0) = u_0(x)$ is given by
\begin{align}\label{ulim}
u(x,t) = 2i\ntlim_{k\to\infty} (kM(x,t,k))_{12},
\end{align}
where $M(x,t,k)$ is the unique solution of the $L^2$-RH problem
\begin{align}\label{RHM}
\begin{cases}
M(x, t, \cdot) \in I + \dot{E}^2(\C \setminus \R),\\
M_+(x,t,k) = M_-(x, t, k) J(x, t, k) \quad \text{for a.e.} \ k \in \R,
\end{cases}
\end{align}
with
\begin{align}\label{Jdef}
J(x,t,k) = \begin{pmatrix} 1 - |r(k)|^2	& -\overline{r(\bar{k})}e^{-2ikx - 8ik^3t}	\\
  r(k) e^{2ikx + 8ik^3t}	& 1 \end{pmatrix}, \qquad k \in \R.
\end{align}
The function $r(k)$ satisfies
\begin{subequations}\label{rsymmbound}
\begin{align}\label{rsymm}
  r(k) = \overline{r(-\bar{k})}, \qquad k \in \R,
\end{align}
and
\begin{align}\label{rbound}
  \sup_{k \in \R} |r(k)| < 1.
\end{align}
\end{subequations}
An elaborate analysis of (\ref{mujdef}) shows that if 
\begin{align}\label{ugjassump}
\begin{cases}
u_0 \in C^{m+1}(\R), &
	\\
(1+|x|)^{n}u_0^{(i)}(x) \in L^1(\R), & i = 0,1, \dots, m+1,
\end{cases}
\end{align}
for some integers $n,m \geq 1$, then $r \in C^n(\R)$ and
\begin{align}
& r^{(j)}(k) = O(k^{-m-1}), \qquad |k| \to \infty, \  k \in \R, \  j = 0, 1, \dots, n.
\end{align}
Moreover, if the initial data $u_0(x)$  satisfy (\ref{ugjassump}) for $n=1$ and $m=4$, then the limit in (\ref{ulim}) exists for each $(x,t) \in \R \times [0, \infty)$ and defines a classical solution $u(x,t)$ of (\ref{mkdv}) with initial data $u(x,0) = u_0(x)$. The above facts can be proved using the inverse scattering approach, see \cite{AC1991, FT2007, BDT1988, IN1986}; in the case of the half-line problem, detailed proofs of the analogous results can be found in \cite{Lmkdvrigorous} (see, in particular, Theorem 7 of \cite{Lmkdvrigorous}).

Since the matrix $\re J = \frac{1}{2}(J + \bar{J}^T)$ is positive definite for $k \in \R$, the homogeneous RH problem determined by $(\R, J)$ has only the zero solution (see Theorem 9.3 of \cite{Z1989}). Hence Lemma \ref{Fredholmzerolemma} implies that the $L^2$-RH problem (\ref{RHM}) has a unique solution whenever $r \in C(\R) \cap L^\infty(\R) \cap L^2(\R)$.

\section{Asymptotics in the similarity sector}\nequation\label{similaritysec}
The goal of this section is to prove Theorem \ref{mainth1} which provides an asymptotic formula for the solution of the mKdV equation in the similarity sector. 
The proof is essentially an application of Theorem \ref{steepestdescentth}.

The similarity sector is the asymptotic region of the $(x,t)$-plane characterized by
\begin{align}\label{similaritysector}
t > 1, \qquad -N t < x < 0, \qquad (-x)^3/t \to \infty,  \qquad \text{$N$ constant},
\end{align}
where the inequality $-Nt < x$ defines the left boundary of the sector while the
condition $(-x)^3/t \to \infty$ defines its right boundary.
Letting $\zeta = x/t < 0$ and defining the variables $k_0 = k_0(\zeta)$ and $\tau = \tau(x,t)$ by
\begin{align}
k_0 = \sqrt{- \frac{\zeta}{12}}, \qquad \tau = 12 t k_0^3 = \frac{|x|^{3/2}}{\sqrt{12 t}},
\end{align}
the condition $(-x)^3/t \to \infty$ can be written as $\tau \to \infty$.

\begin{theorem}\label{mainth1}
Suppose $r(k)$ is a function in $C^{11}(\R)$ which satisfies (\ref{rsymmbound}) and
$$r^{(j)}(k) = O(k^{-4 + 2j}), \qquad |k| \to \infty, \  k \in \R, \  j = 0, 1, 2.$$
Then, for any $\alpha \in [\frac{1}{2},1)$ and $N > 0$, the function $u(x,t)$ defined by (\ref{ulim}) satisfies
\begin{align}\label{ufinal}
u(x,t) = \frac{1}{\sqrt{t k_0}}\bigl[u_a(x,t) + O\bigl(\tau^{-\frac{\alpha}{2}}\bigr)\bigr],\qquad
\tau \to \infty, \  -N t < x < 0,
\end{align}
where the error term is uniform with respect to $x$ in the given range and the function $u_a$ is defined by
\begin{align}\label{uadef}
& u_a(x,t) = \sqrt{\frac{\nu(\zeta)}{3}} \cos\big(16tk_0^3 - \nu(\zeta) \ln(192t k_0^3) + \phi(\zeta)\big)
\end{align}
with
\begin{align}\label{phizetadef}
& \phi(\zeta) = \arg \Gamma(i\nu(\zeta)) + \frac{\pi}{4} - \arg r(k_0) 
- \frac{1}{\pi} \int_{-k_0}^{k_0}  \ln\bigg(\frac{1- |r(s)|^2}{1- |r(k_0)|^2}\bigg) \frac{ds}{s - k_0},
	\\ \label{nudef}
& \nu(\zeta) = -\frac{1}{2\pi} \ln(1 - |r(k_0)|^2).
\end{align}
\end{theorem}

\begin{remark}\label{decayremark}\upshape
Theorem \ref{mainth1} applies whenever the function $r(k)$ satisfies the stated assumptions, regardless of whether this function is defined via (\ref{rdef}) or not. If we assume that $r(k)$ is defined in terms of $u_0(x)$ via (\ref{rdef}), we can ensure that the function $u(x,t)$ defined in (\ref{ulim}) constitutes a solution of (\ref{mkdv}) with initial data $u_0(x)$. More precisely, recalling the remarks at the end of Section \ref{mkdvsec}, we obtain the following corollary of Theorem \ref{mainth1}: Suppose $u_0$ satisfies (\ref{ugjassump}) with $n = 11$ and $m = 4$, i.e.,
$$u_0 \in C^5(\R) \qquad \text{and} \qquad (1+|x|)^{11}u_0^{(i)}(x) \in L^1(\R), \quad i = 0,1, \dots, 5.$$
Then the function $u(x,t)$  defined by (\ref{ulim}) is a well-defined classical ($C^3$ in $x$ and $C^1$ in $t$) solution of (\ref{mkdv}) with initial data $u(x,0) = u_0(x)$, and the asymptotics of $u(x,t)$ in the similarity sector is given by (\ref{ufinal}). 
\end{remark}

\begin{remark}\upshape
The conclusion of Theorem \ref{mainth1} can be stated more explicitly as follows: Given any $\alpha \in [\frac{1}{2},1)$ and $N> 0$, there exist constants $T>0$ and $K>0$ such that the limit in (\ref{ulim}) exists and the function $u(x,t)$ defined by (\ref{ulim}) satisfies
  $$\bigg|u(x,t) - \frac{u_a(x,t)}{\sqrt{t k_0}}\bigg| \leq \frac{K}{\tau^{\frac{\alpha}{2}}\sqrt{t k_0}}$$
 whenever $(x,t) \in \R \times [0,\infty)$ satisfy $-N t < x < 0$ and $\tau > T$.
\end{remark}

\begin{remark}\upshape
The asymptotic regime considered in Theorem \ref{mainth1} is denoted by II in \cite{DZ1993}.
Our expression for the leading asymptotics (\ref{uadef}) coincides with that given in \cite{DZ1993}, except that the expression for $\phi(\zeta)$ in \cite{DZ1993} contains $-\pi/4$ instead of $+\pi/4$. The discrepancy arises because our spectral function $r(k)$ is related to the spectral function $r_{DZ}(k)$ of \cite{DZ1993} by $r(k) = ir_{DZ}(k)$. 
\end{remark}

\begin{remark}\upshape
Theorem \ref{mainth1} applies to a more general class of solutions than the analogous result in \cite{DZ1993}. The formulas of  \cite{DZ1993} were established under the assumption that the initial data $u_0(x)$ belong to the Schwartz class $\mathcal{S}(\R)$ of rapidly decreasing functions and this assumption implies that $r(k)$ also belongs to the Schwartz class $\mathcal{S}(\R)$.
\end{remark}

\begin{remark}\upshape
Theorem \ref{mainth1} is stated under the assumption that $r \in C^{11}(\R)$. We will use this regularity of $r(k)$ in order to establish appropriate analytic approximations of the functions $r_2$ and $r_3$ defined in equation (\ref{r1234def}). Indeed, the proof of Lemma \ref{decompositionlemma2} relies on the fact that the function $F$ defined in (\ref{Fdef}) lies in $H^2(\R)$, which in turn depends on the existence of the Taylor series in (\ref{r3taylor}). The proof of the analogous analytic approximations of the functions $r_1$ and $r_4$ also defined in (\ref{r1234def}) only requires that $r \in C^7(\R)$. 
\end{remark}

\subsection{Proof of Theorem \ref{mainth1}}
\begin{figure}
\begin{center}
\bigskip
 \begin{overpic}[width=.6\textwidth]{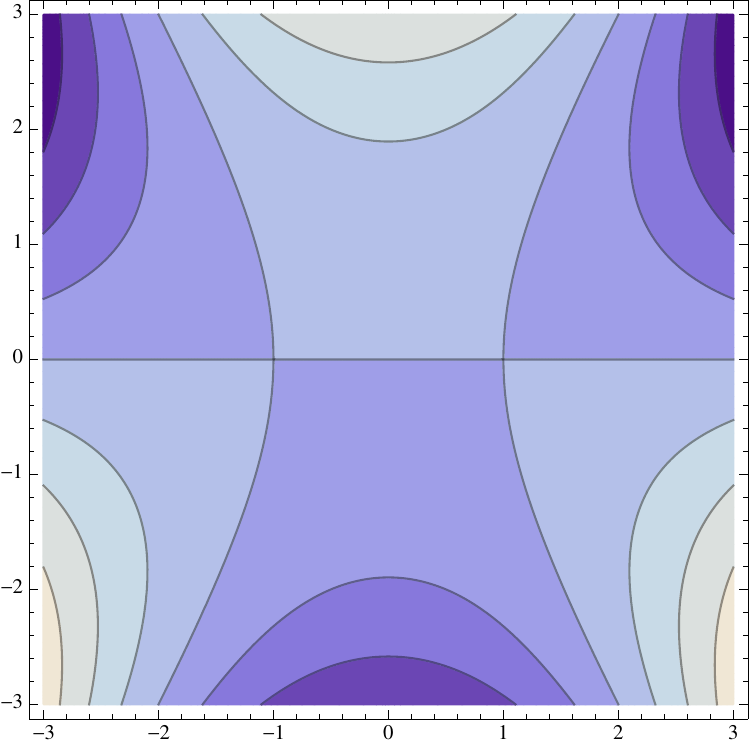}
  \put(62,47){$k_0$}
  \put(35,47){$-k_0$}
  \put(14,42){$\re \Phi > 0$}
  \put(43,67){$\re \Phi > 0$}
  \put(72,42){$\re \Phi > 0$}
  \put(14,58){$\re \Phi < 0$}
  \put(43,35){$\re \Phi < 0$}
  \put(72,58){$\re \Phi < 0$}
    \end{overpic}
   \bigskip
   \begin{figuretext}\label{rePhi.pdf}
      Contour plot of $\re \Phi(\zeta, k)$ in the complex $k$-plane for $k_0 = 1$. Light (dark) regions correspond to positive (negative) values of $\re \Phi$. 
            \end{figuretext}
   \end{center}
\end{figure}

Let $N > 0$ be given and let $\mathcal{I}$ denote the interval $\mathcal{I} = [-N,0)$. 
Let $M(x, t, \cdot) \in I + \dot{E}^2(\C \setminus \R)$ denote the unique solution of the RH problem (\ref{RHM}). The jump matrix $J$ defined in (\ref{Jdef}) involves the exponentials $e^{\pm t\Phi(\zeta,k)}$ where
$$\Phi(\zeta, k) = 2ik\zeta + 8ik^3.$$
It follows that there are two stationary points located at the points where $\frac{d\Phi}{dk} = 0$, i.e. at $k = \pm k_0$. The real part of $\Phi$ is shown in Figure \ref{rePhi.pdf}.
In order to apply the steepest descent result of Theorem \ref{steepestdescentth}, we need to transform the RH problem in such a way that the jump matrix has decay everywhere as $t \to \infty$ except near the two stationary points. This can be achieved by performing an appropriate triangular factorization of the jump matrix followed by a contour deformation. 
For $|k|>k_0$, it is easy to achieve an appropriate factorization. By conjugating the RH problem (\ref{RHM}), we can achieve an appropriate factorization also for $|k| < k_0$.

\medskip
{\bf Step 1: Conjugate.}
Let
$$\Delta(\zeta, k) = \begin{pmatrix} \delta(\zeta, k)^{-1} & 0 \\  0 & \delta(\zeta, k)   \end{pmatrix},$$
where
\begin{align}\label{deltadef}
 \delta(\zeta, k) = e^{\frac{1}{2\pi i} \int_{-k_0}^{k_0} \ln(1- |r(s)|^2) \frac{ds}{s - k}}, \qquad  k \in \C \setminus \R.
\end{align}
The function $\delta$ satisfies the following jump condition across the real axis:
\begin{align}\label{deltajump}
 \delta_+(\zeta, k) = \begin{cases} 
\delta_-(\zeta, k), & |k| > k_0, \\
\delta_-(\zeta, k)(1 - |r(k)|^2), & |k| < k_0, 
\end{cases} \quad k \in \R.
\end{align}
Moreover, the symmetry (\ref{rsymm}) implies that
\begin{align}\label{chideltasymm}
\delta(\zeta, k) = \overline{\delta(\zeta, -\bar{k})} =  \delta(\zeta, -k)^{-1}.
\end{align}
It follows that $\Delta$  obeys the symmetries
\begin{align}\label{Deltasymm}
  \Delta(\zeta, k) = \overline{\Delta(\zeta, -\bar{k})} =  \Delta(\zeta, -k)^{-1}.
\end{align}

\begin{lemma}\label{Deltalemma}
The $2 \times 2$-matrix valued function $\Delta(\zeta, k)$ satisfies
$$\Delta(\zeta, \cdot)^{\pm 1} \in I + \dot{E}^2(\C \setminus \R) \cap E^\infty(\C \setminus \R),$$
for each $\zeta \in \mathcal{I}$.
\end{lemma}
\begin{proof}
First note that
\begin{align}\label{deltarep}
\delta(\zeta, k) = \bigg(\frac{k-k_0}{k + k_0}\bigg)^{i\nu} e^{\chi(\zeta, k)},
\end{align}
where $\nu(\zeta)$ is given by (\ref{nudef}),
\begin{align}\label{chidef}
\chi(\zeta, k) = \frac{1}{2\pi i}\int_\R \psi(\zeta, s) \frac{ds}{s - k},
\end{align}
and the function $\psi(\zeta,s)$ is defined by 
$$\psi(\zeta, s) = \begin{cases} \ln\big(\frac{1- |r(s)|^2}{1- |r(k_0)|^2}\big), & -k_0 < s < k_0,
\\
0, & \text{otherwise}.
\end{cases}$$
Since $\psi(\zeta, \cdot) \in H^1(\R)$, we have $\chi(\zeta, \cdot) \in E^\infty(\C \setminus \R)$ for each $\zeta \in \mathcal{I}$; in fact, see Lemma 23.3 in \cite{BDT1988},
\begin{align}\label{chibound}
\sup_{\zeta \in \mathcal{I}} \sup_{k \in \C \setminus \R} |\chi(\zeta, k)| < \sup_{\zeta \in \mathcal{I}} \|\psi(\zeta, \cdot)\|_{H^1(\R)} < \infty.
\end{align}
Hence $\delta(\zeta, \cdot), \delta(\zeta, \cdot)^{-1} \in E^\infty(\C \setminus \R)$ for each $\zeta \in \mathcal{I}$.

Since $\chi(\zeta, k) = O(k^{-1})$ uniformly as $k \to \infty$, Lemma \ref{EpCnlemma} shows that $\delta(\zeta, \cdot)^{\pm 1} \in 1 + \dot{E}^2(\C \setminus \R)$. 
\end{proof}

By Lemma \ref{Deltalemma} and Lemma \ref{deformationlemma}, the function $\tilde{M}$ defined by
$$\tilde{M}(x,t,k) = M(x,t,k)\Delta(\zeta, k)$$
satisfies the $L^2$-RH problem
$$\begin{cases}
\tilde{M}(x, t, \cdot)  \in I + \dot{E}^2(\C \setminus \R),\\
\tilde{M}_+(x,t,k) = \tilde{M}_-(x, t, k) \tilde{J}(x, t, k)  \quad \text{for a.e.} \ k \in \R,
\end{cases}$$
where
\begin{align} \nonumber
\tilde{J}&(x,t,k) = \Delta_-^{-1}(\zeta, k) J(x,t,k) \Delta_+(\zeta, k)
	\\ \label{tildeJdef}
& = \begin{pmatrix} \frac{\delta_-(\zeta, k)}{\delta_+(\zeta, k)}(1 - |r(k)|^2)	& -\delta_-(\zeta, k)\delta_+(\zeta, k)  \overline{r(\bar{k})}e^{-t\Phi(\zeta,k)}	\\
  \frac{1}{\delta_-(\zeta, k)\delta_+(\zeta, k)} r(k) e^{t\Phi(\zeta,k)}	& \frac{\delta_+(\zeta, k)}{\delta_-(\zeta, k)} \end{pmatrix},  \qquad k \in \R.
\end{align}
In view of the jump (\ref{deltajump}) of $\delta(\zeta, k)$, this gives, for $k \in \R$,
\begin{align*}
\tilde{J}(x,t,k) = 
\begin{cases}
 \begin{pmatrix} 1 - |r(k)|^2	& - \delta(\zeta, k)^2 \overline{r(\bar{k})} e^{-t\Phi(\zeta,k)}		\\
 \delta(\zeta, k)^{-2} r(k) e^{t\Phi(\zeta,k)}	& 1 \end{pmatrix}, 	& |k| > k_0,	\\
 \begin{pmatrix} 1 & -\delta_+(\zeta, k)^2 \frac{\overline{r(\bar{k})}}{1 - |r(k)|^2} e^{-t\Phi(\zeta,k)}	\\
    \delta_-(\zeta, k)^{-2} \frac{r(k)}{1 - |r(k)|^2} e^{t\Phi(\zeta,k)}	& 1 - |r(k)|^2 \end{pmatrix},   & |k| < k_0. 
\end{cases}
\end{align*}

The upshot of the above conjugation is that we can now factorize the jump matrix as follows:
\begin{align}\label{tildeJonR}
  \tilde{J} = 
\begin{cases}
 B_u^{-1} B_l,   & |k| > k_0, \  k \in \R, \\
 b_l^{-1}b_u 	& |k| < k_0,  \  k \in \R, 
\end{cases}
\end{align}
where
\begin{align}\nonumber
& B_l =  \begin{pmatrix} 1 & 0	\\
  \delta(\zeta, k)^{-2} r_1(k) e^{t\Phi(\zeta,k)}	& 1 \end{pmatrix}, 
	 \qquad
b_u = \begin{pmatrix} 1 & -\delta_+(\zeta, k)^2 r_2(k) e^{-t\Phi(\zeta,k)}	\\
0 & 1  \end{pmatrix}, 
	\\ \label{Bbdef}
&b_l = \begin{pmatrix} 1 &0 \\
 -  \delta_-(\zeta, k)^{-2} r_3(k) e^{t\Phi(\zeta,k)} & 1 \end{pmatrix}, 
 	\qquad
B_u =  \begin{pmatrix} 1	& \delta(\zeta, k)^2 r_4(k) e^{-t\Phi(\zeta,k)}	\\
0	& 1 \end{pmatrix},
\end{align}
and the functions $\{r_j(k)\}_1^4$ are defined by
\begin{align}\nonumber
& r_1(k) = r(k), \qquad
r_2(k) = \frac{\overline{r(\bar{k})}}{1 - r(k)\overline{r(\bar{k})}},
	\\ \label{r1234def}
& r_3(k) = \frac{r(k)}{1 - r(k)\overline{r(\bar{k})}}, \qquad
r_4(k) = \overline{r(\bar{k})}.	
\end{align}
Our next goal is to deform the contour. However, we first need to introduce analytic approximations of $\{r_j(k)\}_1^4$.

\begin{figure}
\begin{center}
\bigskip
 \begin{overpic}[width=.5\textwidth]{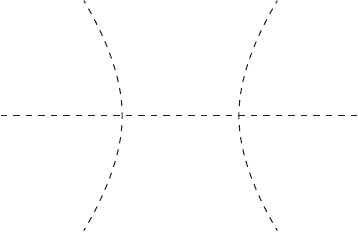}
 \put(85,46){$U_1$}
 \put(48,50){$U_2$}
 \put(48,12){$U_3$}
 \put(85,16){$U_4$}
 \put(10,46){$U_1$}
  \put(10,16){$U_4$}
 \put(23.5,27){$-k_0$}
 \put(69,27){$k_0$}
 \end{overpic}
   \bigskip
   \begin{figuretext}\label{Omegas2.pdf}
      The domains $\{U_j\}_1^4$ in the complex $k$-plane. The dashed curves are the curves on which $\re \Phi = 0$.
      \end{figuretext}
   \end{center}
\end{figure}

\medskip
{\bf Step 2: Introduce analytic approximations.}
The following lemma describes how to decompose $r_j$, $j = 1, \dots, 4$, into an analytic part $r_{j,a}$ and a small remainder $r_{j,r}$.
We introduce open domains $U_j = U_j(\zeta)$, $j = 1, \dots, 4$, as in Figure \ref{Omegas2.pdf} so that 
\begin{align}\label{Ujdef}
\{k \, | \, \re \Phi(\zeta, k) < 0\} = U_1 \cup U_3, \qquad \{k \, | \, \re \Phi(\zeta, k) > 0\} = U_2 \cup U_4.
\end{align}

\begin{lemma}\label{decompositionlemma2}
There exist decompositions
\begin{align*}
r_j(k) =
\begin{cases}
 r_{j,a}(x, t, k) + r_{j,r}(x, t, k), \qquad j = 1,4, \  |k| > k_0,  \  k \in \R,
	\\
 r_{j,a}(x, t, k) + r_{j,r}(x, t, k), \qquad j = 2,3, \  |k| < k_0,  \  k \in \R,
\end{cases}
\end{align*}
where the functions $\{r_{j,a}, r_{j,r}\}_{j=1}^4$ have the following properties:
\begin{enumerate}[$(a)$]
\item For each $\zeta \in \mathcal{I}$ and each $t > 0$, $r_{j,a}(x, t, k)$ is defined and continuous for $k \in \bar{U}_j$ and analytic for $k \in U_j$, $j = 1, \dots, 4$.

\item The functions $\{r_{j,a}\}_1^4$ satisfy, for each $K > 0$,
\begin{align}\nonumber
& |r_{j, a}(x, t, k) - r_j(k_0)| \leq 
C_K |k - k_0| e^{\frac{t}{4}|\re \Phi(\zeta,k)|}, 
	\\ \label{rjaestimatea}
& \hspace{4cm} k \in \bar{U}_j, \  |k| \leq K, \  \zeta \in \mathcal{I}, \  t > 0,  \  j = 1, \dots, 4, 
\end{align}
where the constant $C_K$ is independent of $\zeta, t, k$ but may depend on $K$.

\item The functions $r_{1,a}$ and $r_{4,a}$ satisfy
\begin{align}\label{rjaestimateb}
& |r_{j, a}(x, t, k)| \leq \frac{C}{1 + |k|} e^{\frac{t}{4}|\re \Phi(\zeta,k)|}, \qquad  k \in \bar{U}_j, \  \zeta \in \mathcal{I}, \  t > 0, \  j = 1, 4, 
\end{align}
where the constant $C$ is independent of $\zeta, t, k$.

\item The $L^1, L^2$, and $L^\infty$ norms on $(-\infty, -k_0) \cup (k_0, \infty)$ of the functions $r_{1,r}(x, t, \cdot)$ and $r_{4,r}(x, t, \cdot)$ are $O(t^{-3/2})$ as $t \to \infty$ uniformly with respect to $\zeta \in \mathcal{I}$.

\item The $L^1, L^2$, and $L^\infty$ norms on $(-k_0, k_0)$ of the functions $r_{2,r}(x, t, \cdot)$ and $r_{3,r}(x, t, \cdot)$ are $O(t^{-3/2})$ as $t \to \infty$ uniformly with respect to $\zeta \in \mathcal{I}$.

\item The following symmetries are valid:
\begin{align}\label{rsymmetries}
r_{j,a}(\zeta, t, k) = \overline{r_{j,a}(\zeta, t, -\bar{k})}, \quad
r_{j,r}(\zeta, t, k) = \overline{r_{j,r}(\zeta, t, -\bar{k})}, \qquad j = 1, \dots, 4.
\end{align}
\end{enumerate}
\end{lemma}
\begin{proof}
We will derive decompositions of $r_1(k)$ and $r_3(k)$; the decompositions of $r_2(k)$ and $r_4(k)$ can be obtained from these by Schwartz conjugation.

\smallskip
{\bf Decomposition of $r_1(k)$.}
Let $U_1 = U_1^+ \cup U_1^-$, where $U_1^+$ and $U_1^-$ denote the parts of $U_1$ in the right and left half-planes respectively. We will derive a decomposition of $r_1$ in $U_1^+$ and then extend it to $U_1^-$ by symmetry. 

Since $r \in C^{11}(\R)$, we have
\begin{align*}
 r_1^{(n)}(k) = \frac{d^n}{dk^n}\bigg(\sum_{j=0}^6 p_j(\zeta) (k - k_0)^j\bigg) + O((k-k_0)^{7-n}), \qquad k \to k_0, \  k \in \R, \  n = 0,1,2,
\end{align*}
where the functions $\{p_j(\zeta)\}_0^6$ are defined by $p_j(\zeta) := r_1^{(j)}(k_0)/j!$. 
We let
$$f_0(\zeta, k) = \sum_{j=4}^{10} \frac{a_j(\zeta)}{(k - i)^j},$$
where $\{a_j(\zeta)\}_4^{10}$ are such that
\begin{align}\label{linearconditions2}
& f_0(\zeta, k) = \sum_{j=0}^6 p_j(\zeta) (k-k_0)^j + O((k-k_0)^7), \qquad k \to k_0,
\end{align}
for each $\zeta \in \mathcal{I}$. It is easy to verify that (\ref{linearconditions2}) imposes seven linear conditions on the $a_j(\zeta)$'s that are linearly independent for each $\zeta \in \mathcal{I}$; hence the coefficients $a_j(\zeta)$ exist and are unique. The $a_j(\zeta)$'s are polynomials in $\{p_j(\zeta)\}_0^6$ with coefficients that are polynomials in $k_0$. Thus, since the interval $\mathcal{I}$ is bounded, we have $\sup_{\zeta \in \mathcal{I}} |a_j(\zeta)| < \infty$ for each $j$.

Let $f(\zeta,k) = r_1(k) - f_0(\zeta,k)$. The following properties hold:
\begin{enumerate}[$(i)$]
\item For each $\zeta \in \mathcal{I}$, $f_0(\zeta, k)$ is a rational function of $k \in \C$ with no poles in $\bar{U}_1^+$.

\item $f_0(\zeta, k)$ coincides with $r_1(k)$ to order six at $k_0$ and to order three at $\infty$; more precisely
\begin{align}\label{fcoincide2}
 \frac{\partial^n f}{\partial k^n} (\zeta, k) =
\begin{cases}
 O((k-k_0)^{7 - n}), & k \to k_0, 
	\\
O(k^{-4+2n}), & k \to \infty, 
 \end{cases}
 \quad  k \in \R, \  \zeta \in \mathcal{I}, \  n = 0,1,2,
\end{align}
where the error terms are uniform with respect to $\zeta \in \mathcal{I}$.
\end{enumerate}

\begin{figure}
\begin{center}
\bigskip\medskip
 \begin{overpic}[width=.50\textwidth]{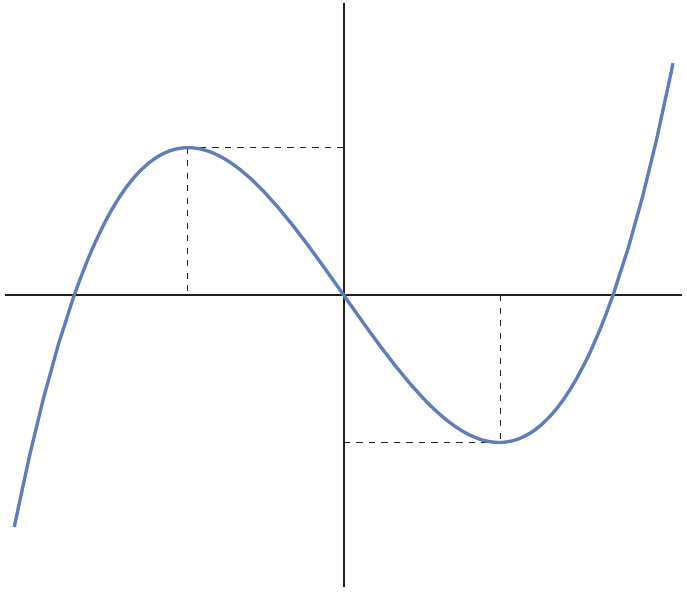}
      \put(102,42){$k$}
      \put(48.7,89.5){$\phi$}
      \put(71,46){$k_0$}
      \put(22,37){$-k_0$}
      \put(52,63){$16k_0^3$}
      \put(34,21){$-16k_0^3$}
       \end{overpic}
     \begin{figuretext}\label{phiofk.pdf}
       Graph of the function $k \mapsto \phi(\zeta, k)$ defined in (\ref{phidef}).
     \end{figuretext}
     \end{center}
\end{figure}

The decomposition of $r_1(k)$ can now be derived as follows.
For each  $\zeta \in \mathcal{I}$, the map $k \mapsto \phi = \phi(\zeta, k)$ where
\begin{align}\label{phidef}
  \phi = -i\Phi(\zeta, k) = -24 k_0^2 k + 8k^3.
\end{align}  
is a bijection  $[k_0, \infty) \to [-16 k_0^3, \infty)$ (see Figure \ref{phiofk.pdf}), so we may define a function $F(\zeta, \phi)$ by
\begin{align}\label{Fdef2}
F(\zeta, \phi) = \begin{cases} \frac{(k-i)^3}{k-k_0} f(\zeta, k), &  \phi \geq -16k_0^3, \\
0, & \phi < -16k_0^3,
\end{cases} \quad \zeta \in \mathcal{I}, \  \phi \in \R.
\end{align}
For each $\zeta \in \mathcal{I}$, the function $F(\zeta, \phi)$ is $C^{11}$ in $\phi$ for $\phi \neq -16k_0^3$ and 
\begin{align}\label{dnFdphin}
\frac{\partial^n F}{\partial \phi^n}(\zeta, \phi) = \bigg(\frac{1}{24 (k^2 -k_0^2)} \frac{\partial }{\partial k}\bigg)^n 
\bigg[\frac{(k-i)^3}{k-k_0} f(\zeta, k)\bigg], \qquad \phi \geq -16k_0^3.
\end{align}
Equations (\ref{fcoincide2}) and (\ref{dnFdphin}) together with the trivial inequalities
\begin{align}\label{kk0ineq}
\frac{k}{k + k_0} \leq 1 \quad \text{and} \quad \frac{k- k_0}{k + k_0} \leq 1 \quad \text{for}\quad k\geq k_0,
\end{align}
show that $F(\zeta, \cdot) \in C^1(\R)$ for each $\zeta$ and that
\begin{align*}
\bigg| \frac{\partial^n F}{\partial \phi^n}(\zeta, \phi)\bigg| \leq 
 \frac{C}{(1 + |\phi|)^{\frac{2}{3}}}, \qquad \phi \in (-16k_0^3, \infty), \  \zeta \in \mathcal{I}, \  n = 0,1,2.
\end{align*}
Hence
\begin{align}\label{supdFdphi}
\sup_{\zeta \in \mathcal{I}} \bigg\|\frac{\partial^n F}{\partial \phi^n}(\zeta, \cdot)\bigg\|_{L^2(\R)} < \infty, \qquad n = 0,1,2.
\end{align}
In particular, $F(\zeta, \cdot)$ belongs to the Sobolev space $H^2(\R)$ for each $\zeta \in \mathcal{I}$.
We conclude that the Fourier transform $\hat{F}(\zeta, s)$ defined by
\begin{align}\label{Fhatdef}
\hat{F}(\zeta, s) = \frac{1}{2\pi} \int_{\R} F(\zeta, \phi) e^{-i\phi s} d\phi,
\end{align}
satisfies
\begin{align}\label{FFhat}
F(\zeta, \phi) =  \int_{\R} \hat{F}(\zeta, s) e^{i\phi s} ds
\end{align}
and
\begin{align}\label{x2Fhat}
\sup_{\zeta \in \mathcal{I}} \|s^2 \hat{F}(\zeta, s)\|_{L^2(\R)} < \infty.
\end{align}
Equations (\ref{Fdef2}) and (\ref{FFhat}) imply
$$ \frac{k-k_0}{(k-i)^3}\int_{\R} \hat{F}(\zeta, s) e^{s\Phi(\zeta,k)} ds 
= \begin{cases} f(\zeta, k), &  k \geq k_0, \\
0, & k < k_0, 
 \end{cases} \qquad \zeta \in \mathcal{I}.$$
Writing
$$f(\zeta, k) = f_a(x, t, k) + f_r(x, t, k), \qquad \zeta \in \mathcal{I}, \  t > 0, \  k \geq k_0,$$
where the functions $f_a$ and $f_r$ are defined by
\begin{align*}
& f_a(x,t,k) = \frac{k-k_0}{(k-i)^3}\int_{-\frac{t}{4}}^\infty \hat{F}(\zeta,s) e^{s\Phi(\zeta,k)} ds, \qquad \zeta \in \mathcal{I}, \  t > 0, \  k \in \bar{U}_1^+,  
	\\
& f_r(x,t,k) = \frac{k-k_0}{(k-i)^3}\int_{-\infty}^{-\frac{t}{4}} \hat{F}(\zeta,s) e^{s\Phi(\zeta,k)} ds,\qquad \zeta \in \mathcal{I}, \  t > 0, \   k \geq k_0,
\end{align*}
we infer that $f_a(x, t, \cdot)$ is continuous in $\bar{U}_1^+$ and analytic in $U_1^+$. 
Furthermore,
\begin{align}\nonumber
 |f_a(x, t, k)| 
&\leq \frac{|k-k_0|}{|k - i|^3} \|\hat{F}(\zeta,\cdot)\|_{L^1(\R)}  \sup_{s \geq -\frac{t}{4}} e^{s \re \Phi(\zeta,k)}
\leq \frac{C|k-k_0|}{|k - i|^3}  e^{\frac{t}{4} |\re \Phi(\zeta,k)|} 
	\\ \label{faest}
& \hspace{5cm} \zeta \in \mathcal{I}, \  t > 0, \  k \in \bar{U}_1^+,
\end{align}
and
\begin{align}\nonumber
|f_r(x, t, k)| & \leq \frac{|k-k_0|}{|k - i|^3}  \int_{-\infty}^{-\frac{t}{4}} s^2 |\hat{F}(\zeta,s)| s^{-2} ds
 \leq \frac{C}{1 + |k|^2}  \| s^2 \hat{F}(\zeta,s)\|_{L^2(\R)} \sqrt{\int_{-\infty}^{-\frac{t}{4}} s^{-4} ds}  
 	\\ \label{frest}
&  \leq \frac{C}{1 + |k|^2} t^{-3/2}, \qquad \zeta \in \mathcal{I}, \  t > 0, \  k \geq k_0.
\end{align}
Hence the $L^1$, $L^2$, and $L^\infty$ norms of $f_r$ on $(k_0, \infty)$ are $O(t^{-3/2})$ uniformly with respect to $\zeta \in \mathcal{I}$. Letting
\begin{align*}
& r_{1,a}(x, t, k) = f_0(\zeta, k) + f_a(x, t, k), \qquad k \in \bar{U}_1^+,
	\\
& r_{1,r}(x, t, k) = f_r(x, t, k), \qquad k \geq k_0.
\end{align*}
we find a decomposition of $r_1$ for $k > k_0$ with the properties listed in the statement of the lemma. We use the symmetry (\ref{rsymmetries}) to extend this decomposition to $k < -k_0$.

\smallskip
{\bf Decomposition of $r_3$.}
Following \cite{DZ1993}, we split $r_3(k)$ into even and odd parts as follows:
$$r_3(k) = r_+(k^2) + k r_-(k^2), \qquad k \in \R,$$
where $r_\pm:[0,\infty) \to \C$ are defined by
$$r_+(s) = \frac{r_3(\sqrt{s}) + r_3(-\sqrt{s})}{2}, \qquad 
r_-(s) = \frac{r_3(\sqrt{s}) - r_3(-\sqrt{s})}{2\sqrt{s}}, \qquad s \geq 0.$$
The symmetry $r_3(k) = \overline{r_3(-\bar{k})}$ implies
$$r_+(s)  = \re r_3(\sqrt{s}), \qquad r_-(s) = \frac{i\im r_3(\sqrt{s})}{\sqrt{s}}, \qquad s \geq 0.$$
Since $r$, and hence also  $r_3$, belongs to $C^{11}(\R)$, we can write the Taylor series 
\begin{align}\label{r3taylor}
r_3(k) = \sum_{j=0}^{10} q_j k^j + \frac{1}{10!} \int_0^k r_3^{(11)}(t) (k-t)^{10} dt,
\end{align}
where $q_j := r_3^{(j)}(0)/j!$. 
It follows that
\begin{subequations}\label{rero}
\begin{align}
& r_+(s) = \sum_{i=0}^{5} q_{2i} s^i + \frac{1}{2\cdot 10!} \int_0^{\sqrt{s}} (r_3^{(11)}(t) - r_3^{(11)}(-t))(\sqrt{s} - t)^{10} dt,
	\\
& r_-(s) = \sum_{i=0}^{4} q_{2i + 1} s^i + \frac{1}{2\cdot 10! \sqrt{s}} \int_0^{\sqrt{s}} (r_3^{(11)}(t) + r_3^{(11)}(-t))(\sqrt{s} - t)^{10} dt.
\end{align}
\end{subequations}
Since $r_3 \in C^{11}(\R)$, the equations (\ref{rero}) show that the derivatives $r_\pm^{(j)}(s)$ are bounded on $[0, K]$ for $0 \leq j \leq 5$ for each $K > 0$.
Letting $\{p_j^\pm(\zeta)\}_0^4$ denote the coefficients of the Taylor series representations
\begin{align*}
& r_\pm(k^2) = \sum_{j=0}^{4} p_j^\pm(\zeta) (k^2 - k_0^2)^j + \frac{1}{4!}\int_{k_0^2}^{k^2} r_\pm^{(5)}(t)(k^2 - t)^4 dt,
\end{align*}
we infer that the function $f_0(\zeta,k)$ defined by 
$$f_0(\zeta, k) =  \sum_{j=0}^{4} p_j^+(\zeta) (k^2 - k_0^2)^j + k \sum_{j=0}^{4} p_j^-(\zeta) (k^2 - k_0^2)^j$$
has the following properties:
\begin{enumerate}[$(i)$]
\item $f_0(\zeta, k)$ is a polynomial in $k \in \C$ whose coefficients are bounded functions of $\zeta \in \mathcal{I}$.

\item The difference $f(\zeta,k) = r_3(k) - f_0(\zeta,k)$ satisfies 
\begin{align}\label{fcoincide3}
 \frac{\partial^n f}{\partial k^n} (\zeta, k) \leq C |k^2-k_0^2|^{5-n}, \qquad \zeta \in \mathcal{I}, \  -k_0 \leq k \leq k_0, \  n = 0,1,2,
\end{align}
where $C$ is independent of $\zeta$ and $k$. 

\item $f_0(\zeta, k) = \overline{f_0(\zeta, -\bar{k})}$ for $k \in \C$.
\end{enumerate}

The decomposition of $r_3(k)$ can now be derived as follows. The function $k \mapsto \phi$ defined in (\ref{phidef}) is a bijection  $[-k_0,k_0] \to [-16k_0^3,16k_0^3]$ (see Figure \ref{phiofk.pdf}), so we may define 
a function $F(\zeta, \phi)$ by 
\begin{align}\label{Fdef}
F(\zeta, \phi) = \begin{cases} \frac{1}{k^2-k_0^2} f(\zeta, k), & |\phi| \leq 16 k_0^3, \\
0, & |\phi| > 16 k_0^3,
\end{cases} \quad \zeta \in \mathcal{I}, \  \phi \in \R.
\end{align}
For each $\zeta \in \mathcal{I}$, the function $F(\zeta, \phi)$ is $C^{11}$ in $\phi$ for $\phi \neq \pm 16k_0^3$ and 
\begin{align}\label{dnFdphin2}
\frac{\partial^n F}{\partial \phi^n}(\zeta, \phi) = \bigg(\frac{1}{24 (k^2 -k_0^2)} \frac{\partial }{\partial k}\bigg)^n\frac{f(\zeta, k)}{k^2-k_0^2}, \qquad |\phi| \leq 16 k_0^3.
\end{align}
Equations (\ref{fcoincide3}) and (\ref{dnFdphin2}) show that $F(\zeta, \cdot) \in C^1(\R)$ for each $\zeta$ and that
\begin{align*}
\bigg| \frac{\partial^n F}{\partial \phi^n}(\zeta, \phi) \bigg| \leq C, \qquad |\phi| \leq 16 k_0^3, \  \phi \in \R, \  n = 0,1,2.
\end{align*}
Hence $F$ satisfies (\ref{supdFdphi}) and the Fourier transform $\hat{F}(\zeta, s)$ defined by (\ref{Fhatdef}) satisfies (\ref{FFhat}) and (\ref{x2Fhat}).
Equations (\ref{FFhat}) and (\ref{Fdef}) imply
$$(k^2-k_0^2)\int_{\R} \hat{F}(\zeta, s) e^{s\Phi(\zeta,k)} ds = \begin{cases} f(\zeta, k), &  |k| \leq k_0, \\
0, & |k|> k_0, \end{cases} \qquad \zeta \in \mathcal{I}, \  k \in \R.$$
Writing
$$f(\zeta, k) = f_a(x, t, k) + f_r(x, t, k), \qquad \zeta \in \mathcal{I}, \  t > 0, \  -k_0 \leq k \leq k_0,$$
where the functions $f_a$ and $f_r$ are defined by
\begin{align*}
& f_a(x,t,k) = (k^2-k_0^2)\int_{-\frac{t}{4}}^\infty \hat{F}(\zeta,s) e^{s\Phi(\zeta,k)} ds, \qquad \zeta \in \mathcal{I}, \  t > 0, \  k \in \bar{U}_3,  
	\\
& f_r(x,t,k) = (k^2-k_0^2)\int_{-\infty}^{-\frac{t}{4}} \hat{F}(\zeta,s) e^{s\Phi(\zeta,k)} ds,\qquad \zeta \in \mathcal{I}, \  t > 0, \   -k_0 \leq k \leq k_0,
\end{align*}
we infer that $f_a(x, t, \cdot)$ is continuous in $\bar{U}_3$ and analytic in $U_3$. 
Estimating $f_a$ and $f_r$ as in  (\ref{faest}) and (\ref{frest}), we find
\begin{align*}
&  |f_a(\zeta, t, k)| \leq C |k^2-k_0^2| e^{\frac{t}{4} |\re \Phi(\zeta,k)|}, \qquad \zeta \in \mathcal{I}, \  t > 0, \  k \in \bar{U}_3,
	\\
& |f_r(\zeta, t, k)| \leq C t^{-3/2}, \qquad \zeta \in \mathcal{I}, \  t > 0, \   -k_0 \leq k \leq k_0.
\end{align*}
It follows that  
$$r_{3,a}(x, t, k) = f_0(\zeta, k) + f_a(x, t, k), \qquad r_{3,r}(x, t, k) = f_r(x, t, k),$$
provides a decomposition of $r_3$ with the properties listed in the statement of the lemma.
The symmetries (\ref{rsymmetries}) are satisfied since $F(\zeta, \phi) = \overline{F(\zeta, -\phi)}$ for $\phi \in \R$.
\end{proof}

\begin{figure}
\begin{center}
\bigskip
 \begin{overpic}[width=.65\textwidth]{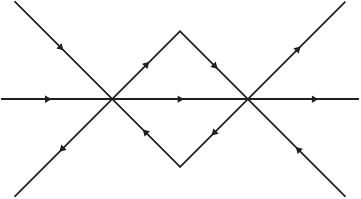}
      \put(67.5,22){\small $k_0$}
      \put(27.5,22){\small $-k_0$}
      \put(85,34){\small $V_1$}
      \put(48,51){\small $V_2$}
      \put(48, 34){\small $V_3$}
      \put(48,18){\small $V_4$}
      \put(48,2){\small $V_5$}
      \put(85,18){\small $V_6$}
      \put(10,34){\small $V_1$}
      \put(10,18){\small $V_6$}
      \put(10,18){\small $V_6$}
      \put(102,26.5){\small $\Gamma$}
       \end{overpic}
    \bigskip
     \begin{figuretext}\label{GammaEfig}
       The jump contour  $\Gamma$ and the open sets $\{V_j\}_1^{6}$.
     \end{figuretext}
     \end{center}
\end{figure}

\medskip
{\bf Step 3: Deform.}
Let $\Gamma$ be the contour consisting of $\R$ together with the four half-lines 
$$\big\{k_0 + u e^{\pm \frac{i\pi}{4}} \, |  -\sqrt{2} k_0 < u < \infty\big\}, \qquad
\big\{-k_0 + u e^{\pm \frac{3i\pi}{4}} \, | -\sqrt{2} k_0 < u < \infty\big\},$$
oriented as in Figure \ref{GammaEfig}. Let $\{V_j\}_1^{6}$ be the open sets shown in Figure \ref{GammaEfig}. Write
$$B_l = B_{l,r}B_{l,a}, \qquad
b_u = b_{u,r}b_{u,a}, \qquad
b_l = b_{l,r}b_{l,a}, \qquad
B_u = B_{u,r}B_{u,a},$$
where $\{B_{l,a}, b_{u,a}, b_{l,a}, B_{u,a}\}$ and $\{B_{l,r}, b_{u,r}, b_{l,r}, B_{u,r}\}$ denote the matrices $\{B_{l}, b_{u}, b_{l}, B_{u}\}$ with $\{r_j(k)\}_1^4$ replaced with $\{r_{j,a}(k)\}_1^4$ and $\{r_{j,r}(k)\}_1^4$, respectively.
The estimates (\ref{rjaestimatea}) and (\ref{rjaestimateb}) together with Lemma \ref{EpCnlemma} imply that 
\begin{align*}
\begin{cases}
B_{l,a}(x,t,\cdot)^{\pm 1} \in I + \dot{E}^2(V_1) \cap E^\infty(V_1),
	\\
b_{u,a}(x,t,\cdot)^{\pm 1} \in I + \dot{E}^2(V_3) \cap E^\infty(V_3),
	\\
b_{l,a}(x,t,\cdot)^{\pm 1}  \in I + \dot{E}^2(V_4) \cap E^\infty(V_4),
	\\
B_{u,a}(x,t,\cdot)^{\pm 1} \in I + \dot{E}^2(V_6) \cap E^\infty(V_6),
\end{cases}
\end{align*}
for each $\zeta \in \mathcal{I}$ and each $t > 0$.
Hence, we may apply Lemma \ref{deformationlemma} to deduce that the function $m(x,t,k)$ defined by
\begin{align}\label{betweendef}
m(x,t,k) = \begin{cases}  
\tilde{M}(x,t,k) B_{l,a}(x,t,k)^{-1}, & k \in V_1, \\
\tilde{M}(x,t,k) b_{u,a}(x,t,k)^{-1}, & k \in V_3, \\
\tilde{M}(x,t,k) b_{l,a}(x,t,k)^{-1}, & k \in V_4, \\
\tilde{M}(x,t,k) B_{u,a}(x,t,k)^{-1}, & k \in V_6, \\
\tilde{M}(x,t,k), & \text{elsewhere},
\end{cases}
\end{align}
satisfies the $L^2$-RH problem
\begin{align}\label{RHm2}
\begin{cases} m(x, t, \cdot) \in I + \dot{E}^2(\hat{\C} \setminus \Gamma), \\
m_+(x, t, k) = m_-(x, t, k) v(x, t, k) \quad \text{for a.e.} \ k \in \Gamma, 
\end{cases} 
\end{align}
where, in view of (\ref{tildeJdef}) and (\ref{tildeJonR}), the jump matrix $v$ is given by
\begin{align}\label{vdef}
v = \begin{cases}
B_{l,a} = \begin{pmatrix} 1 & 0	\\
  \delta(\zeta, k)^{-2} r_{1,a}(x,t,k) e^{t\Phi(\zeta,k)}	& 1 \end{pmatrix}, & k \in \bar{V}_1 \cap \bar{V}_2, \\
b_{u,a} = \begin{pmatrix} 1 & -\delta(\zeta, k)^2 r_{2,a}(x,t,k) e^{-t\Phi(\zeta,k)}	\\
0 & 1  \end{pmatrix}, & k \in \bar{V}_2 \cap \bar{V}_3, \\
b_{l,a} = \begin{pmatrix} 1 &0 \\
 -  \delta(\zeta, k)^{-2} r_{3,a}(x,t,k) e^{t\Phi(\zeta,k)} & 1 \end{pmatrix}, & k \in \bar{V}_4 \cap \bar{V}_5,
 	\\
B_{u,a} = \begin{pmatrix} 1	& \delta(\zeta, k)^2 r_{4,a}(x,t,k) e^{-t\Phi(\zeta,k)}	\\
0	& 1 \end{pmatrix}, & k \in \bar{V}_5 \cap \bar{V}_6.
	\\
B_{u,r}^{-1} B_{l,r}, & k \in \bar{V}_1 \cap \bar{V}_6, 
	\\
b_{l,r}^{-1}b_{u,r}, & k \in \bar{V}_3 \cap \bar{V}_4.
\end{cases}
\end{align}
 
From the symmetries (\ref{chideltasymm}) and (\ref{rsymmetries}), we infer that $v$ satisfies
\begin{align}\label{vsymm2}
v(x,t,k) = \overline{v(x, t, -\bar{k})}, \qquad k \in \Gamma.
\end{align}

\medskip
{\bf Step 4: Apply Theorem \ref{steepestdescentth}.}
We verify that Theorem \ref{steepestdescentth} can be applied to the interval $\mathcal{I} = [-N,0)$, the contour $\Gamma$, and the jump matrix $v$ with 
\begin{align} \nonumber
& \epsilon = \frac{k_0}{2}, \qquad
\rho = \epsilon \sqrt{-i \frac{\partial^2 \Phi}{\partial k^2}(\zeta, k_0)}
 = \epsilon \sqrt{48 k_0}, \qquad \tau = t\rho^2 = 12 k_0^3 t,
 	\\ \nonumber
&  q(\zeta) = e^{-2\chi(\zeta,k_0)} r(k_0) e^{2i\nu(\zeta) \ln(2\sqrt{48}k_0^{3/2})}, \qquad \nu(\zeta) = -\frac{1}{2\pi} \ln(1 -  |r(k_0)|^2),
	\\  \label{choices}
& \phi(\zeta, z) = \Phi\biggl(\zeta, k_0 + \frac{\epsilon}{\rho}z\biggr)
 = -16i k_0^3 + \frac{iz^2}{2} + \frac{i z^3}{12 \rho}.
\end{align}

\begin{figure}
\begin{center}
\bigskip
 \begin{overpic}[width=.48\textwidth]{Gammasteep.pdf}
     \put(48,50){$\Gamma$}
           \put(67,22){\tiny $k_0$}
      \put(27.2,22){\tiny $-k_0$}
   \end{overpic}
       \quad 
 \begin{overpic}[width=.481\textwidth]{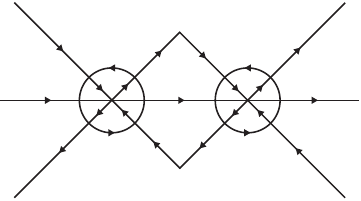}
      \put(48,50){$\hat{\Gamma}$}
            \put(67,22){\tiny $k_0$}
      \put(27.2,21){\tiny $-k_0$}
      \end{overpic}
     \begin{figuretext}\label{Gammafig}
       The contours $\Gamma$ and $\hat{\Gamma}$.
     \end{figuretext}
     \end{center}
\end{figure}

The contours $\Gamma$ and $\hat{\Gamma}$ are shown in Figure \ref{Gammafig}. The conditions ($\Gamma$1)-($\Gamma$4) are clearly satisfied. 
Since the contour $k_0^{-1}\hat{\Gamma}$ is independent of $\zeta$, a scaling argument shows that $\|\mathcal{S}_{\hat{\Gamma}}\|_{\mathcal{B}(L^2(\hat{\Gamma}))}$ is independent of $\zeta$. In particular, $\mathcal{S}_{\hat{\Gamma}}$ is uniformly bounded on $L^2(\hat{\Gamma})$.
Equation (\ref{winL1L2Linf}) follows from (\ref{vdef}) and the estimates in Lemma \ref{decompositionlemma2}.
Clearly $\det v = 1$. The symmetry condition (\ref{vsymm}) follows from (\ref{vsymm2}).

We next verify (\ref{wL12infty}).
Let $w = v - I$ and let $\Gamma'$ denote the contour obtained from $\Gamma$  by removing the crosses $\pm k_0 + X^\epsilon$. By parts $(d)$ and $(e)$ of Lemma \ref{decompositionlemma2}, the $L^1$ and $L^\infty$ norms of $w$ are $O(t^{-\frac{3}{2}})$ on $\R$ uniformly with respect to $\zeta \in \mathcal{I}$.
Let $\gamma$ denote the part of $\Gamma'$ that belongs to the line $k_0 + \R e^{\frac{i\pi}{4}}$, i.e.
$$\gamma = \bigg\{k_0 + ue^{\frac{i\pi}{4}} \, \bigg| \, u \in \bigg(-\sqrt{2}k_0, -\frac{k_0}{2}\bigg] \cup \bigg[\frac{k_0}{2}, \infty\bigg) \bigg\}.$$ 
Let $k = k_0 + ue^{\frac{i\pi}{4}}$. Then
\begin{align}\label{rePhiest}
\re \Phi(\zeta, k) = -4u^2(6k_0 + \sqrt{2} u) < - 16 u^2 k_0 \qquad \text{for} \quad -\sqrt{2}k_0 < u < \infty.
\end{align}
Together with (\ref{rjaestimateb}) this yields
\begin{align*}
|r_{1,a}(x, t, k) e^{t \Phi(\zeta, k)}| & \leq Ce^{-\frac{3t}{4}|\re \Phi(\zeta,k)|}
 \leq Ce^{-12 tu^2k_0} 
 \leq \frac{C t u^2k_0 e^{-12 tu^2k_0}}{t u^2 k_0} 
 	\\
& 
\leq \frac{C}{t u^2 k_0}
\leq \frac{Ck_0^2}{u^2 \tau}, \qquad \frac{k_0}{2}< u < \infty.
\end{align*}
Similarly, by (\ref{rjaestimatea}),
\begin{align*}
|r_{3,a}(x, t, k) e^{t \Phi(\zeta, k)}| & \leq C e^{-\frac{3t}{4}|\re \Phi(\zeta,k)|}
\leq \frac{Ck_0^2}{u^2 \tau}, \qquad -\sqrt{2}k_0 < u < -\frac{k_0}{2}.
\end{align*}
Hence
\begin{align*}
& \|w\|_{L^1(\gamma)} \leq  \frac{Ck_0^2}{\tau} \bigg(\int_{-\sqrt{2}k_0}^{-\frac{k_0}{2}} + \int_{\frac{k_0}{2}}^\infty\bigg)  u^{-2} du= O(k_0 \tau^{-1}),
	\\
& \|w\|_{L^\infty(\gamma)} = O(\tau^{-1}).
\end{align*}
This shows that the estimates in (\ref{wL12infty}) hold also on $\gamma$. Since similar arguments apply to the remaining parts of $\Gamma'$, this  verifies (\ref{wL12infty}).

Equation (\ref{vdef}) implies that (\ref{vjdef}) and (\ref{smallcrossjump}) are satisfied with
\begin{align*}
\begin{cases}
R_1(\zeta, t, z) =  \delta(\zeta,k)^{-2} r_{1,a}(x,t,k)z^{2i\nu(\zeta)},
	\\
R_2(\zeta, t, z) = \delta(\zeta, k)^2 r_{2,a}(x,t,k) z^{-2i\nu(\zeta)},
	\\
R_3(\zeta, t, z) = \delta(\zeta,k)^{-2} r_{3,a}(x,t,k) z^{2i\nu(\zeta)},
	\\
R_4(\zeta, t, z) = \delta(\zeta,k)^2 r_{4,a}(x,t,k) z^{-2i\nu(\zeta)},
\end{cases}
\end{align*}
where $k$ and $z$ are related by $k = k_0 + \frac{\epsilon z}{\rho}$. 

It is clear that $\phi$ satisfies (\ref{phiassumptions}) and (\ref{Phiz3estimate}). 
The following estimate proves  (\ref{rephiestimateb}):
$$\re \phi(\zeta, z) = \frac{|z|^2}{2}\bigg(1 \pm \frac{|z|}{6\sqrt{2} \rho}\bigg) \geq \frac{|z|^2}{4}, \qquad z \in X_2^{\rho} \cup X_4^{\rho}, \  \zeta \in \mathcal{I},$$
where the plus and minus signs are valid for $z \in X_4^{\rho}$ and $z \in X_2^{\rho}$ respectively. The proof of (\ref{rephiestimatea}) is similar.
Since $|q(\zeta)| = |r(k_0)|$, we have $\sup_{\zeta \in \mathcal{I}} |q(\zeta)| < 1$ and $\nu(\zeta) = -\frac{1}{2\pi} \ln(1 - |q(\zeta)|^2)$.

Finally, we show that given any $\alpha \in [1/2, 1)$, there exists an $L >0$ such that the inequalities (\ref{Lipschitzconditions}) hold. Let  $k = k_0 + \frac{\epsilon z}{\rho}$.
Using the expression (\ref{deltarep}) for $\delta(\zeta, k)$, we may write
$$R_1(\zeta, t, z) = e^{-2\chi(\zeta, k)} r_{1,a}(x,t,k) e^{2i\nu(\zeta) \ln((k + k_0)\sqrt{48k_0})} , \qquad z \in X_1.$$ 
Thus,
$$R_1(\zeta, t, 0) = e^{-2\chi(\zeta, k_0)} r_{1,a}(x,t,k_0)  e^{2i\nu(\zeta)\ln(2\sqrt{48}k_0^{3/2})}.$$
Now $r_{1,a}(x,t,k_0) = r(k_0)$ by (\ref{rjaestimatea}). Hence $R_1(\zeta, t, 0) = q(\zeta)$.
Similarly, we find 
$$R_2(\zeta, t, 0) = \frac{\overline{q(\zeta)}}{1 - |q(\zeta)|^2}, \qquad
R_3(\zeta, t, 0) = \frac{q(\zeta)}{1 - |q(\zeta)|^2}, \qquad
R_4(\zeta, t, 0) = \overline{q(\zeta)}.$$
We establish the estimate (\ref{Lipschitzconditions}) in the case of $z \in X_1^{\rho}$; the other cases are similar.
Note that $z \in X_1^{\rho}$ corresponds to $k \in k_0 + X_1^\epsilon$.

We have, for $z \in X_1^\rho$,
\begin{align*}
|R_1(\zeta, t, z) &- q(\zeta)| \leq 
\left|e^{-2\chi(\zeta, k)} - e^{-2\chi(\zeta, k_0)}\right| \left|r_{1,a}(x,t,k) e^{2i\nu(\zeta) \ln((k + k_0)\sqrt{48k_0})}\right|
	\\
& + \left| e^{-2\chi(\zeta,k_0)}\right| \left|r_{1,a}(x,t,k) - r(k_0)\right| \left| e^{2i\nu(\zeta) \ln((k + k_0)\sqrt{48k_0})}\right|
	\\
& + \left| e^{-2\chi(\zeta, k_0)} r(k_0)\right| \left|e^{2i\nu(\zeta)\ln((k + k_0)\sqrt{48k_0})} - e^{2i\nu(\zeta) \ln(2\sqrt{48}k_0^{3/2})}\right|.
\end{align*}
The functions $e^{2i\nu(\zeta)\ln((k + k_0)\sqrt{48k_0})}$ and $e^{-2\chi(\zeta, k_0)}$ are uniformly bounded with respect to $k \in k_0 + X_1^\epsilon$ and $\zeta \in \mathcal{I}$. 
Moreover, employing the estimate 
$$|\re \phi(\zeta, ve^{\frac{i\pi}{4}} )| = \bigg|-\frac{v^2}{2}\bigg(1 + \frac{v}{6\sqrt{2} \rho}\bigg)\bigg|
\leq \frac{2v^2}{3}, \qquad -\rho \leq v \leq \rho,$$
we see that equation (\ref{rjaestimatea}) yields
\begin{align*}
|r_{1, a}(x, t, k) - r(k_0)| 
& \leq C |k -k_0| e^{\frac{t}{4}|\re \Phi(\zeta,k)|} 
 = C \frac{\epsilon|z|}{\rho} e^{\frac{t}{4}|\re \phi(\zeta,z)|} 
 	\\
&  \leq C \frac{\epsilon|z|}{\rho} e^{\frac{t |z|^2}{6}}, \qquad z \in X_1^{\rho}.
\end{align*}
Thus,
\begin{align}\nonumber
 |R_1(\zeta, t, z) - q(\zeta)| \leq &\; C e^{\frac{t |z|^2}{6}}
 \left|e^{-2\chi(\zeta, k)} - e^{-2\chi(\zeta, k_0)}\right| + C \frac{\epsilon |z|}{\rho}e^{\frac{t|z|^2}{6}}
	\\ \label{R1minusq}
& + C \big|1 - e^{-2i\nu(\zeta)\ln(\frac{k + k_0}{2k_0})} \big|, \qquad \zeta \in \mathcal{I}, \  t > 0, \  z \in X_1^{\rho}.
\end{align}
The estimate in (\ref{Lipschitzconditions}) for $z \in X_1^{\rho}$ now follows from the following lemma.

\begin{lemma}
The following inequalities are valid for all $\zeta \in \mathcal{I}$ and all $k \in k_0 + X_1^\epsilon$:
\begin{align}\label{term1}
& \left|e^{-2\chi(\zeta, k)} - e^{-2\chi(\zeta, k_0)}\right| \leq C |k-k_0| (1 + |\ln|k-k_0||), 
	\\ \label{term3}
& \left|1 - e^{-2i\nu(\zeta)\ln(\frac{k + k_0}{2k_0})}\right|  \leq C k_0^{-1} |k-k_0|,
\end{align}
where the constant $C$ is independent of $\zeta$ and $k$. 
\end{lemma}
\proofbegin
We first prove that
\begin{align}\label{chiminuschi}
|\chi(\zeta, k) - \chi(\zeta, k_0)| \leq C |k-k_0|\big(1 + |\ln|k-k_0|| \big), \qquad \zeta \in \mathcal{I}, \  k \in k_0 + X_1^\epsilon.
\end{align}
Integration by parts in the definition (\ref{chidef}) of $\chi$ gives 
\begin{align}\label{chizetal}
 \chi(\zeta, k) = - \frac{1}{2\pi i} \int_{-k_0}^{k_0} \ln(s-k) d\ln\bigl(1- |r(s)|^2\bigr).
\end{align}
Hence
$$|\chi(\zeta, k) - \chi(\zeta, k_0)| 
= \frac{1}{2\pi} \biggl| \int_{-k_0}^{k_0} \ln\biggl(\frac{s - k}{s - k_0}\biggr) d\ln\bigl(1- |r(s)|^2\bigr) \biggr|.$$
This gives
\begin{align}\label{chichichi}
& |\chi(\zeta, k) - \chi(\zeta, k_0)| 
 \leq C \int_{-k_0}^{k_0} \biggl| \ln\frac{s - k}{s - k_0}\biggr| ds
\leq C \int_0^{2k_0} \biggl| \ln\biggl(1 + \frac{k - k_0}{u}\biggr) \biggr|du.
\end{align}
The change of variables $v =|k - k_0|/u$ yields
\begin{align*}
\int_0^{2k_0} \biggl| \ln\biggl(1 + \frac{k - k_0}{u}\biggr) \biggr|du
= |k-k_0| \int_{\frac{|k-k_0|}{2k_0}}^\infty \bigl| \ln(1 + v e^{\frac{i\pi}{4}}) \bigr| \frac{dv}{v^2}.
\end{align*}
Since 
$$\bigl| \ln(1 + v e^{\frac{i\pi}{4}}) \bigr|  \leq C \begin{cases}
v, & 0 \leq v \leq 2, 
	\\
\ln v, & 2 \leq v < \infty,
\end{cases}$$
we conclude that
\begin{align*}
 |\chi(\zeta, k) - \chi(\zeta, k_0)|
 & \leq C|k - k_0|\bigg(\int_{\frac{|k-k_0|}{2k_0}}^2 \frac{dv}{v} + \int_2^\infty \frac{\ln v}{v^2}dv\bigg)
	\\
& \leq C |k-k_0| (|\ln|k-k_0|| + C), \qquad k \in k_0 + X_1^\epsilon.
\end{align*}
This proves (\ref{chiminuschi}). 

Using the inequality (\ref{ewminus1estimate}) together with (\ref{chibound}) and (\ref{chiminuschi}), we estimate
\begin{align*}
|e^{-2\chi(\zeta, k)} - e^{-2\chi(\zeta, k_0)}|
& \leq |e^{-2\chi(\zeta, k_0)}| |e^{-2[\chi(\zeta, k) - \chi(\zeta, k_0)]} - 1 |
	\\
& \leq C |\chi(\zeta, k) - \chi(\zeta, k_0)| e^{2|\re(\chi(\zeta, k) - \chi(\zeta, k_0))|}
	\\
& \leq C |k-k_0| (1+ |\ln|k-k_0||), \qquad \zeta \in \mathcal{I}, \  k \in k_0 + X_1^\epsilon.
\end{align*}
This proves (\ref{term1}).

By (\ref{ewminus1estimate}),
\begin{align*}
\left|1 - e^{-2i\nu(\zeta)\ln(\frac{k + k_0}{2k_0})}\right| & \leq \bigg|2\nu(\zeta)\ln\bigg(\frac{k + k_0}{2k_0}\bigg)\bigg| e^{|\re(2i\nu(\zeta)\ln(\frac{k + k_0}{2k_0}))|}
	\\
& \leq C \bigg|\ln\biggl(1 + \frac{k-k_0}{2k_0}\biggr)\bigg| \leq C k_0^{-1} |k-k_0|, \qquad \zeta \in \mathcal{I},
\ k \in k_0 + X_1^\epsilon, 
\end{align*}
which proves (\ref{term3}).

\medskip
{\bf Step 5: Find asymptotics.}
Theorem \ref{steepestdescentth} implies that the limit (\ref{ulim}) defining $u(x,t)$ exists for all sufficiently large $\tau$ and is given by
\begin{align*}
u(x,t) & = 2i\ntlim_{k\to\infty} (kM(x,t,k))_{12} = 2i \ntlim_{k\to\infty} (km(x,t,k))_{12}.
\end{align*}
Equation (\ref{limlm12}) of Theorem \ref{steepestdescentth} then yields
\begin{align*}
u(x,t) & = \frac{4 \epsilon \re \beta(\zeta, t)}{\sqrt{\tau}} + O\bigl(\epsilon \tau^{-\frac{1+\alpha}{2}}\bigr)
 = \frac{\re \beta(\zeta, t)}{\sqrt{3t k_0}} + O\bigl(\epsilon \tau^{-\frac{1+\alpha}{2}}\bigr),	
\end{align*}
where $\beta(\zeta, t)$ is defined by (\ref{betadef}). We have
$$\re \beta(\zeta,t) = \sqrt{\nu(\zeta)}\cos\bigg(\frac{\pi}{4} - \arg q(\zeta) + \arg \Gamma(i\nu(\zeta))
 + 16 tk_0^3 - \nu(\zeta)\ln t\bigg).$$
where
$$\arg q(\zeta) = 2\nu(\zeta) \ln(2k_0 \sqrt{48k_0}) + \arg{r(k_0)} + \frac{1}{\pi} \int_{-k_0}^{k_0}  \ln\bigg(\frac{1- |r(s)|^2}{1- |r(k_0)|^2}\bigg) \frac{ds}{s - k_0}.$$
Thus,
\begin{align*}
u(x,t) & = \sqrt{\frac{\nu(\zeta)}{3tk_0}} \cos\big(16tk_0^3 - \nu(\zeta) \ln(192tk_0^3) + \phi(\zeta)\big) + O\bigl(\epsilon \tau^{-\frac{1+\alpha}{2}}\bigr),
\end{align*}
where $\phi(\zeta)$ is defined by (\ref{phizetadef}). This proves (\ref{ufinal}) and completes the proof of Theorem \ref{mainth1}.

\appendix
\section{$L^2$-Riemann-Hilbert problems} \label{RHapp}
\renewcommand{\theequation}{A.\arabic{equation}}\nequation
A theory of $L^p$-RH problems with jumps across Carleson contours was presented in \cite{LCarleson}; here we collect a number of relevant definitions and results. In the context of smooth contours, more information on $L^p$-RH problems can be found in \cite{D1999, DZ2002b, FIKN2006, Z1989}.

Let $\mathcal{J}$ denote the collection of all subsets $\Gamma$ of the Riemann sphere $\hat{\C} = \C \cup \{\infty\}$ such that $\Gamma$ is homeomorphic to the unit circle and
\begin{align}\label{carlesondef}
 \sup_{z \in \Gamma \cap \C} \sup_{r > 0} \frac{|\Gamma \cap D(z, r)|}{r} < \infty,
\end{align}
where $D(z, r)$ denotes the disk of radius $r$ centered at $z$. Curves satisfying (\ref{carlesondef}) are called Carleson curves. Let $1 \leq p < \infty$. If $D$ is the bounded component of $\hat{\C} \setminus \Gamma$ where $\Gamma \in \mathcal{J}$ and $\infty \notin \Gamma$, then a function $f$ analytic in $D$ belongs to the Smirnoff class $E^p(D)$ if there exists a sequence of rectifiable Jordan curves $\{C_n\}_1^\infty$ in $D$, tending to the boundary in the sense that $C_n$ eventually surrounds each compact subdomain of $D$, such that
\begin{align}\label{Epsup}
\sup_{n \geq 1} \int_{C_n} |f(z)|^p |dz| < \infty.
\end{align}
If $D$ is a subset of $\hat{\C}$ bounded by an arbitrary curve in $\mathcal{J}$, $E^p(D)$ is defined as the set of functions $f$ analytic in $D$ for which $f \circ \varphi^{-1} \in E^p(\varphi(D))$, where $\varphi(z) = \frac{1}{z - z_0}$ and $z_0$ is any point in $\C \setminus \bar{D}$. The subspace of $E^p(D)$ consisting of all functions $f \in E^p(D)$ such that $z f(z) \in E^p(D)$ is denoted by $\dot{E}^p(D)$. If $D = D_1 \cup \cdots \cup D_n$ is the union of a finite number of disjoint subsets of $\hat{\C}$ each of which is bounded by a curve in $\mathcal{J}$, then $E^p(D)$ and $\dot{E}^p(D)$ denote the set of functions $f$ analytic in $D$ such that $f|_{D_j} \in E^p(D_j)$ and $f|_{D_j} \in \dot{E}^p(D_j)$ for each $j$, respectively.
We let $E^\infty(D)$ denote the space of bounded analytic functions in $D$. 
A {\it Carleson jump contour} is a connected subset $\Gamma$ of $\hat{\C}$ such that:
\begin{enumerate}[$(a)$]
\item $\Gamma \cap \C$ is the union of finitely many oriented arcs\footnote{A subset $\Gamma \subset \C$ is an {\it arc} if it is homeomorphic to a connected subset $I$ of the real line which contains at least two distinct points.} each pair of which have at most endpoints in common.

\item $\hat{\C} \setminus \Gamma$ is the union of two disjoint open sets $D_+$ and $D_-$ each of which has a finite number of simply connected components in $\hat{\C}$.

\item $\Gamma$ is the positively oriented boundary of $D_+$ and the negatively oriented boundary of $D_-$, i.e. $\Gamma = \partial D_+ = -\partial D_-$.

\item If $\{D_j^+\}_1^n$ and $\{D_j^-\}_1^m$ are the components of $D_+$ and $D_-$, then $\partial D_j^+ \in \mathcal{J}$ for $j = 1, \dots, n$, and $\partial D_j^- \in \mathcal{J}$ for $j = 1, \dots, m$.
\end{enumerate}

Let $n \geq 1$ be an integer and let $\Gamma$ be a Carleson jump contour. Given an $n \times n$-matrix valued function $v: \Gamma \to GL(n, \C)$, a {\it solution of the $L^p$-RH problem determined by $(\Gamma, v)$} is an $n \times n$-matrix valued function $m \in I + \dot{E}^p(\hat{\C} \setminus \Gamma)$ such that the nontangential boundary values $m_\pm$ satisfy $m_+ = m_- v$ a.e. on $\Gamma$. 
If $\tilde{\Gamma}$ denotes the Carleson jump contour $\Gamma$ with the orientation reversed on a subset $\Gamma_0 \subset \Gamma$ and $\tilde{v}$ is defined by $\tilde{v} =v$ on $\Gamma \setminus \Gamma_0$ and $\tilde{v} = v^{-1}$ on $\Gamma_0$,
then we say that $m \in I + \dot{E}^p(D)$ satisfies the $L^p$-RH problem determined by $(\tilde{\Gamma}, \tilde{v})$ if and only if $m$ satisfies the $L^p$-RH problem determined by $(\Gamma, v)$. 

We next list some facts about Smirnoff classes and $L^2$-RH problems; detailed proofs can be found in \cite{LCarleson}. We assume that $\Gamma = \partial D_+ = -\partial D_-$ is a Carleson jump contour.

\subsection{Basic facts}
If $f \in \dot{E}^2(D_+)$ or  $f \in \dot{E}^2(D_-)$, then the nontangential limits of $f(z)$ as $z$ approaches the boundary exist a.e. on $\Gamma$ and the boundary function belongs to $L^2(\Gamma)$.
If $h \in L^2(\Gamma)$, then the Cauchy transform $\mathcal{C}h$ defined by
\begin{align}\label{Cauchytransform}
(\mathcal{C}h)(z) = \frac{1}{2\pi i} \int_\Gamma \frac{h(s)}{s - z} ds, \qquad z \in \C \setminus \Gamma,
\end{align}
satisfies $\mathcal{C}h \in \dot{E}^2(D_+ \cup D_-)$. 
We denote the nontangential boundary values of $\mathcal{C}f$ from the left and right sides of $\Gamma$ by $\mathcal{C}_+ f$ and $\mathcal{C}_-f$ respectively. Then $\mathcal{C}_+$ and $\mathcal{C}_-$ are bounded operators on $L^2(\Gamma)$ and $\mathcal{C}_+ - \mathcal{C}_- = I$.

Let $L^2(\Gamma) + L^\infty(\Gamma)$ denote the space of all functions which can be written as the sum of a function in $L^2(\Gamma)$ and a function in $L^\infty(\Gamma)$. 
Given two functions $w^\pm \in L^2(\Gamma) \cap L^\infty(\Gamma)$, we define the operator $\mathcal{C}_{w}: L^2(\Gamma) + L^\infty(\Gamma) \to L^2(\Gamma)$ by 
\begin{align}\label{Cwdef}
\mathcal{C}_{w}(f) = \mathcal{C}_+(f w^-) + \mathcal{C}_-(f w^+).
\end{align}
Then
\begin{align}\label{Cwnorm}
\|\mathcal{C}_w\|_{\mathcal{B}(L^2(\Gamma))} \leq C \max\big\{\|w^+\|_{L^\infty(\Gamma)}, \|w^-\|_{L^\infty(\Gamma)} \big\}.
\end{align}
where $C = \max\{\|\mathcal{C}_+\|_{\mathcal{B}(L^2(\Gamma))}, \|\mathcal{C}_-\|_{\mathcal{B}(L^2(\Gamma))}\} < \infty$ and $\mathcal{B}(L^2(\Gamma))$ denotes the Banach space of bounded linear maps $L^2(\Gamma) \to L^2(\Gamma)$.

The next lemma shows that if $v = (v^-)^{-1}v^+$ and $w^\pm = \pm v^\pm \mp I$ then the $L^2$-RH problem determined by $(\Gamma, v)$ is equivalent to the following singular integral equation for $\mu \in I + L^2(\Gamma)$:
\begin{align}\label{rhoeq}
\mu - I = \mathcal{C}_w(\mu)  \quad \text{in}\quad L^2(\Gamma).
\end{align}

\begin{lemma}\label{mulemma}
Given $v^\pm: \Gamma \to GL(n, \C)$, let $v = (v^-)^{-1}v^+$, $w^+ = v^+ - I$, and $w^- = I - v^-$. Suppose $v^\pm, (v^\pm)^{-1} \in I +  L^2(\Gamma) \cap L^\infty(\Gamma)$.
If $m \in I + \dot{E}^2(D)$ satisfies the $L^2$-RH problem determined by $(\Gamma, v)$, then $\mu = m_+ (v^+)^{-1} = m_- (v^-)^{-1} \in I + L^2(\Gamma)$ satisfies (\ref{rhoeq}). 
Conversely, if $\mu \in I + L^2(\Gamma)$ satisfies (\ref{rhoeq}), then
$m = I + \mathcal{C}(\mu(w^+ + w^-)) \in I + \dot{E}^2(D)$ satisfies the $L^2$-RH problem determined by $(\Gamma, v)$. 
\end{lemma}

\begin{lemma}\label{Fredholmzerolemma}
Given $v^\pm: \Gamma \to GL(n, \C)$, let $v = (v^-)^{-1}v^+$, $w^+ = v^+ - I$, and $w^- = I - v^-$. Suppose $v^\pm, (v^\pm)^{-1} \in  I + L^2(\Gamma) \cap L^\infty(\Gamma)$ and $v^\pm \in C(\Gamma)$.
If $w^\pm$ are nilpotent matrices, then each of the following four statements implies the other three:
\begin{enumerate}[$(a)$]
\item The map $I - \mathcal{C}_w:L^2(\Gamma) \to L^2(\Gamma)$ is bijective.

\item The $L^2$-RH problem determined by $(\Gamma, v)$ has a unique solution. 

\item The homogeneous $L^2$-RH problem determined by $(\Gamma, v)$ has only the zero solution.

\item  The map $I - \mathcal{C}_w: L^2(\Gamma) \to L^2(\Gamma)$ is injective.
\end{enumerate}
\end{lemma}

\begin{lemma}[Uniqueness]\label{uniquelemma}
Suppose $v: \Gamma \to GL(2, \C)$ satisfies $\det v = 1$ a.e. on $\Gamma$.
If the solution of the $L^2$-RH problem determined by $(\Gamma, v)$ exists, then it is unique and has unit determinant.
\end{lemma}

\begin{lemma}\label{EpCnlemma}
Let $D$ be a subset of $\hat{\C}$ bounded by a curve $\gamma \in \mathcal{J}$. Let $f:D \to \C$ be an analytic function. 
Suppose there exist curves $\{C_n\}_1^\infty \subset \mathcal{J}$ in $D$, tending to $\gamma$ in the sense that $C_n$ eventually surrounds each compact subset of $D \subset \hat{\C}$, such that $\sup_{n \geq 1} \|f(z)\|_{L^2(C_n)} < \infty$.
Then $f \in \dot{E}^2(D)$.
\end{lemma}

\begin{figure}
\begin{center}
 \begin{overpic}[width=.27\textwidth]{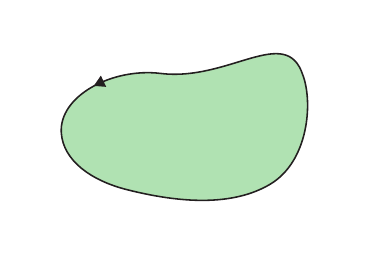}
 \put(49,45){\small $B_+$}
 \put(49,74){\small $B_-$}
 \put(20,66){\small $\gamma$}
 \end{overpic}
 \begin{overpic}[width=.32\textwidth]{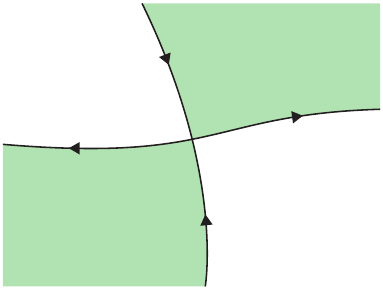}
 \put(70,60){\small $D_+$}
 \put(20,55){\small $D_-$}
 \put(20,16){\small $D_+$}
 \put(70,20){\small $D_-$}
  \put(77,37){\small $\Gamma$}
 \end{overpic}
 \begin{overpic}[width=.32\textwidth]{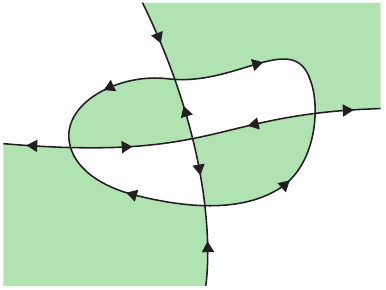}
 \put(84,63){\small $\hat{D}_+$}
 \put(5,63){\small $\hat{D}_-$}
 \put(10,7){\small $\hat{D}_+$}
 \put(84,7){\small $\hat{D}_-$}
 \put(30,43){\small $\hat{D}_+$}
 \put(36,27){\small $\hat{D}_-$}
 \put(60,31){\small $\hat{D}_+$}
 \put(60,48){\small $\hat{D}_-$}
   \put(6,41){\small $\hat{\Gamma}$}
    \put(23,57){$\gamma_+$}
 \put(30,18){\small $\gamma_-$}
 \put(75,21){\small $\gamma_+$}
 \put(65,62){\small $\gamma_-$}
 \end{overpic}
    \bigskip
   \begin{figuretext}\label{gammafig1}
      Examples of contours $\gamma$, $\Gamma$, and $\hat{\Gamma} = \Gamma \cup \gamma$ satisfying the conditions of Lemma \ref{deformationlemma}.
      \end{figuretext}
   \end{center}
\end{figure}

 \begin{lemma}[Contour deformation]\label{deformationlemma}
Let $\gamma \in \mathcal{J}$. Suppose that, reversing the orientation on a subcontour if necessary, $\hat{\Gamma} = \Gamma \cup \gamma$ is a Carleson jump contour, see Figure \ref{gammafig1}. 
Let $B_+$ and $B_-$ be the two components of $\hat{\C} \setminus \gamma$. Let $\hat{D}_\pm$ be the open sets such that $\hat{\C} \setminus \hat{\Gamma} = \hat{D}_+ \cup \hat{D}_-$ and $\partial \hat{D}_+ = - \partial \hat{D}_- = \hat{\Gamma}$. 
Let $\hat{D} = \hat{D}_+ \cup \hat{D}_-$.
Let $\gamma_+$ and $\gamma_-$ be the parts of $\gamma$ that belong to the boundary of $\hat{D}_+ \cap B_+$ and $\hat{D}_- \cap B_+$, respectively. 
Suppose $v: \Gamma \to GL(n, \C)$. Suppose $m_0:\hat{D} \cap B_+ \to GL(n,\C)$ satisfies
\begin{align}\label{m0m0inv}
m_0, m_0^{-1} \in I + \dot{E}^2(\hat{D} \cap B_+) \cap E^\infty(\hat{D} \cap B_+).
\end{align}
Define $\hat{v}:\hat{\Gamma} \to GL(n, \C)$ by
\begin{align*}
\hat{v} 
=  \begin{cases}
 m_{0-} v m_{0+}^{-1} & \text{on} \quad  \Gamma \cap B_+, \\
m_{0+}^{-1} & \text{on} \quad \gamma_+, \\
m_{0-} & \text{on} \quad \gamma_-, \\
v & \text{on} \quad \Gamma \cap B_-.
\end{cases}
\end{align*}
Let $m$ and $\hat{m}$ be related by
\begin{align}\label{hatmdefmm0}
\hat{m} = \begin{cases}
mm_0^{-1} & \text{on} \quad \hat{D} \cap B_+,\\
m & \text{on} \quad \hat{D} \cap B_-.
\end{cases}
\end{align}
Then $m(z)$ satisfies the $L^2$-RH problem determined by $(\Gamma,v)$ if and only if $\hat{m}(z)$ satisfies the $L^2$-RH problem determined by $(\hat{\Gamma}, \hat{v})$.
\end{lemma}

\section{Exact solution in terms of parabolic cylinder functions}\nequation\label{exactapp}
\renewcommand{\theequation}{B.\arabic{equation}}\nequation
Let $X = X_1 \cup \cdots \cup X_4 \subset \C$ be the cross defined in (\ref{Xdef}) and oriented as in Figure \ref{X.pdf}. 
Let $\D \subset \C$ denote the open unit disk and define  the function $\nu:\D \to (0,\infty)$ by 
$\nu(q) = -\frac{1}{2\pi} \ln(1 - |q|^2)$.
We consider the following family of $L^2$-RH problems parametrized by $q \in \D$:
\begin{align}\label{RHmc}
\begin{cases} m^X(q, \cdot) \in I + \dot{E}^2(\C \setminus X), 
	\\
m_+^X(q, z) =  m_-^X(q, z) v^X(q, z) \quad \text{for a.e.} \ z \in X, 
\end{cases} 
\end{align}
where the jump matrix $v^X(q, z)$ is defined by
\begin{align*} 
v^X(q, z) = \begin{cases}
\begin{pmatrix} 1 & 0	\\
  q z^{-2i\nu(q)} e^{\frac{iz^2}{2}}	& 1 \end{pmatrix}, &   z \in X_1, 
  	\\
\begin{pmatrix} 1 & - \frac{\overline{q}}{1 - |q|^2} z^{2i\nu(q)}e^{-\frac{iz^2}{2}}	\\
0 & 1  \end{pmatrix}, &  z \in X_2, 
	\\
\begin{pmatrix} 1 &0 \\
- \frac{q}{1 - |q|^2}z^{-2i\nu(q)} e^{\frac{iz^2}{2}}	& 1 \end{pmatrix}, &  z \in X_3,
	\\
 \begin{pmatrix} 1	& \overline{q} z^{2i\nu(q)}e^{-\frac{iz^2}{2}}	\\
0	& 1 \end{pmatrix}, &  z \in X_4.
\end{cases}
\end{align*}
The matrix $v^X$ has entries that oscillate rapidly as $z \to 0$ and $v^X$ is not continuous at $z = 0$; however $v^X(q, \cdot) - I \in L^2(X) \cap L^\infty(X)$.

The RH problem (\ref{RHmc}) can be solved explicitly in terms of parabolic cylinder functions \cite{I1981}.

\begin{figure}
\begin{center}
\bigskip
 \begin{overpic}[width=.5\textwidth]{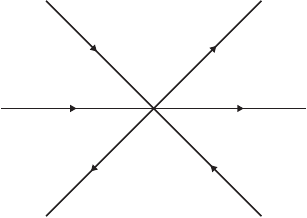}
 \put(48,65){$\Omega_0$}
 \put(48,5){$\Omega_0$}
 \put(77,47){$\Omega_1$}
 \put(17,47){$\Omega_2$}
 \put(17,21){$\Omega_3$}
 \put(77,21){$\Omega_4$}
 \put(102,34){$\R$}
 \put(62,58){$X_1$}
 \put(31,58){$X_2$}
 \put(31,11){$X_3$}
 \put(62,11){$X_4$}
 \end{overpic}
   \bigskip
   \begin{figuretext}\label{Omegas.pdf}
      The sets $\Omega_j$, $j = 1, \dots,4$. 
      \end{figuretext}
   \end{center}
\end{figure}

\begin{theorem}\label{crossth}
  The $L^2$-RH problem (\ref{RHmc}) has a unique solution $m^X(q, z)$ for each $q \in \D$. This solution satisfies
\begin{align}\label{mcasymptotics}
  m^X(q, z) = I + \frac{i}{z}\begin{pmatrix} 0 & -\beta^X(q) \\ \overline{\beta^X(q)} & 0 \end{pmatrix} + O\biggl(\frac{1}{z^2}\biggr), \qquad z \to \infty,  \  q \in \D, 
\end{align}  
where the error term is uniform with respect to $\arg z \in [0, 2\pi]$ and $q$ in compact subsets of $\D$, and the function $\beta^X(q)$ is defined by
\begin{align}\label{betacdef}
\beta^X(q) = \sqrt{\nu(q)} e^{i\left(\frac{\pi}{4} - \arg q + \arg \Gamma(i\nu(q)\right)}, \qquad q \in \D.
\end{align}
Moreover, for each compact subset $K$ of $\D$, 
\begin{align}\label{mXbounded}
\sup_{q \in K} \sup_{z \in \C \setminus X} |m^X(q, z)| < \infty.
\end{align}
\end{theorem}
\proofbegin
Since $\det v^X = 1$, uniqueness follows from Lemma \ref{uniquelemma}.

Define a sectionally analytic function $\tilde{m}^X(q, z)$ by
\begin{align}\label{tildemcexplicit}
  \tilde{m}^X(q, z) = \begin{pmatrix} \psi_{11}(q,z) & \frac{\bigl(\frac{d}{dz} - \frac{iz}{2}\bigr)\psi_{22}(q,z)}{\overline{\beta^X(q)}} \\
\frac{\bigl(\frac{d}{dz} + \frac{iz}{2}\bigr)\psi_{11}(q,z)}{\beta^X(q)}  &  \psi_{22}(q,z) \end{pmatrix}, \qquad q \in \D,\  z \in \C\setminus \R,  
\end{align}
where $\beta^X(q)$ is given by (\ref{betacdef}), the functions $\psi_{11}$ and $\psi_{22}$ are defined by
\begin{align*}
\psi_{11}(q,z) = \begin{cases} e^{-\frac{3\pi \nu(q)}{4}} D_{i\nu(q)}(e^{-\frac{3i\pi}{4}}z), & \im z > 0, \\
e^{\frac{\pi \nu(q)}{4}} D_{i\nu(q)}(e^{\frac{i\pi}{4}}z), & \im z < 0, \end{cases}
	\\
\psi_{22}(q, z) = \begin{cases} e^{\frac{\pi \nu(q)}{4}} D_{-i\nu(q)}(e^{-\frac{i\pi}{4}}z), & \im z > 0, \\
e^{-\frac{3\pi \nu(q)}{4}} D_{-i\nu(q)}(e^{\frac{3i\pi}{4}}z), & \im z < 0, \end{cases}
\end{align*}
and $D_a(z)$ denotes the parabolic cylinder function. Since $D_a(z)$ is an entire function of both $a$ and $z$, $\tilde{m}^X(q, z)$ is analytic in the upper and lower halves of the complex $z$-plane with a jump across the real axis. Observe that $\tilde{m}^X(q, z)$ is regular at $q = 0$ despite the fact that $\beta^X(q)$ vanishes at $q = 0$.

For $j = 1, \dots, 4$, we denote the open domain enclosed by $\R$ and $X_j$ by $\Omega_j$, see Figure \ref{Omegas.pdf}. Let $\Omega_0 = \C \setminus \cup_{j=1}^4 \bar{\Omega}_j$ and define $m^X(q,z)$ by
\begin{align}\label{tildemcmcrelation}
  m^X(q, z) = \tilde{m}^X(q, z) D_j(q, z), \qquad z \in \Omega_j, \  j = 0, \dots, 4,
\end{align}
where
$$D_0(q, z) = \begin{pmatrix} z^{-i\nu(q)} e^{\frac{iz^2}{4}} & 0 \\ 0 & z^{i\nu(q)} e^{-\frac{iz^2}{4}} \end{pmatrix}$$
and
\begin{align*}
& D_1(q,z) = \begin{pmatrix} 1 & 0 \\ -q & 1 \end{pmatrix}D_0(q,z), && 
   D_2(q,z) = \begin{pmatrix} 1 & \frac{\overline{q}}{1 - |q|^2} \\ 0 & 1 \end{pmatrix}D_0(q,z),
  	\\ 
& D_3(q,z) = \begin{pmatrix} 1 & 0 \\ \frac{q}{1 - |q|^2} & 1 \end{pmatrix}D_0(q,z), && 
D_4(q,z) = \begin{pmatrix} 1 & -\overline{q} \\ 0 & 1 \end{pmatrix}D_0(q,z).
\end{align*}

\begin{lemma}\label{lemmaB1}
The function $m^X(q,z)$ is analytic for $z \in \C \setminus X$ and satisfies the jump condition $m_+^X = m_-^X v^X$ a.e. on $X$.
\end{lemma}
\proofbegin
Clearly, the boundary values $m_\pm^X(q, z)$ exist for all $z \in X \setminus \{0\}$.
Since $\tilde{m}^X$ is continuous across  $X$, it is straightforward to verify the jump condition across $X$. For example, the jump of $m^X$ across $X_1$ is given by 
$$(m_-^X)^{-1} m_+^X = D_1^{-1} D_0 = \begin{pmatrix} 1 & 0	\\
  q z^{-2i\nu(q)} e^{\frac{iz^2}{2}}	& 1 \end{pmatrix}, \qquad z \in X_1.$$ 
  
In order to show that $m^X$ is analytic for $z \in \C \setminus X$, it is enough to verify that $m^X$ does not jump across $\R$. We will first prove that $\tilde{m}^X$ satisfies
\begin{align}\label{RHtildemc}
\tilde{m}_+^X(q, z) = \tilde{m}_-^X(q, z) \begin{pmatrix} 1 - |q|^2 & -\overline{q} \\
q & 1 \end{pmatrix}, \qquad q \in \D,\  z \in \R.
\end{align}
Since both $\tilde{m}_+^X(q, z)$ and $\tilde{m}_-^X(q, z)$ satisfy the differential equation 
\begin{align}\label{tildemcODE}
\biggl(\frac{d}{dz} + \frac{iz}{2}\sigma_3\biggr) \tilde{m}^X(q, z) = \begin{pmatrix} 0 & \beta^X(q) \\ \overline{\beta^X(q)} & 0 \end{pmatrix} \tilde{m}^X(q, z),
\end{align}
the jump matrix $\tilde{v}^X(q, z) := \tilde{m}_-^X(q,z)^{-1}\tilde{m}_+^X(q, z)$ is independent of $z$. Evaluating at $z = 0$ and using the identities 
$$\beta^X(q) = \frac{\sqrt{2\pi} e^{\frac{i\pi}{4}} e^{-\frac{\pi \nu(q)}{2}}}{\Gamma(-i\nu(q)) q}, \qquad \beta^X(q) \overline{\beta^X(q)} = \nu(q), \qquad q \in \D,$$ 
and
$$D_a(0) = \frac{2^{\frac{a}{2}} \sqrt{\pi}}{\Gamma(\frac{1-a}{2})}, \qquad
D_a'(0) = - \frac{2^{\frac{1+a}{2}} \sqrt{\pi}}{\Gamma(-\frac{a}{2})},$$
we find the following equation which proves (\ref{RHtildemc}):
\begin{align*}
 \tilde{v}^X&(q, z) = \tilde{v}^X(q,0)
 = \begin{pmatrix} e^{\frac{\pi \nu}{4}} D_{i\nu}(0) & \frac{e^{-\frac{3\pi \nu}{4}}e^{\frac{3i\pi}{4}}D_{-i\nu}'(0)}{\overline{\beta^X(q)}} 	\\
\frac{e^{\frac{\pi \nu}{4}}e^{\frac{i\pi}{4}}D_{i\nu}'(0) }{\beta^X(q)} & e^{-\frac{3\pi \nu}{4}} D_{-i\nu}(0)\end{pmatrix}^{-1}
	\\
& \times \begin{pmatrix} e^{-\frac{3\pi \nu}{4}} D_{i\nu}(0) & \frac{e^{\frac{\pi \nu}{4}}e^{-\frac{i\pi}{4}}D_{-i\nu}'(0)}{\overline{\beta^X(q)}} 	\\
\frac{e^{-\frac{3\pi \nu}{4}}e^{-\frac{3i\pi}{4}}D_{i\nu}'(0) }{\beta^X(q)} & e^{\frac{\pi \nu}{4}} D_{-i\nu}(0)\end{pmatrix}
 = \begin{pmatrix} 1- |q|^2 & -\overline{q} \\ q & 1 \end{pmatrix}, \qquad q \in \D, \  z \in \R.
\end{align*}

Since $z^{i\nu}$ has a branch cut along the negative real axis, the jump of $m^X$ across $\R$ is given by
\begin{align}\label{jumpDDDD}
(m_-^X)^{-1} m_+^X =  \begin{cases} D_4^{-1} \begin{pmatrix} 1 - |q|^2 & -\overline{q} \\
q & 1 \end{pmatrix} D_1, & z > 0, \\
D_{0,-}^{-1}\begin{pmatrix} 1 - |q|^2 & 0 \\
0 & \frac{1}{1 - |q|^2} \end{pmatrix}  D_{0,+}, & z < 0, \end{cases}
\end{align}
where $D_{0,+}$ and $D_{0,-}$ denote the values of $D_0$ for $z$ just above and below the negative real axis respectively:
$$D_{0,\pm}(q,z) =  \begin{pmatrix} e^{-i\nu(q)(\ln|z| \pm i\pi)} e^{\frac{iz^2}{4}} & 0 \\ 0 & e^{i\nu(q)(\ln|z| \pm i\pi)} e^{-\frac{iz^2}{4}} \end{pmatrix}, \qquad z < 0.$$
Simplification of (\ref{jumpDDDD}) shows that $m^X$ has no jump across $\R$.
\proofendcontinue

\begin{lemma}\label{lemmaB2}
 $m^X(q,z)$ satisfies the asymptotic formula (\ref{mcasymptotics}) as $z \to \infty$. 
\end{lemma}
\proofbegin
Let $\delta > 0$ be an arbitrarily small positive constant (unrelated to the function $\delta$ used earlier in the paper).
The parabolic cylinder function satisfies the asymptotic formula \cite{OLBC2010}
\begin{align*}
D_a(z) = & \; z^a e^{-\frac{z^2}{4}}\biggl(1 - \frac{a(a-1)}{2z^2} + O\bigl(z^{-4}\bigr)\biggr)
	\\
& - \frac{\sqrt{2\pi}e^{\frac{z^2}{4}} z^{-a-1}}{\Gamma(-a)} \biggl(1 + \frac{(a+1)(a+2)}{2z^2} + O\bigl(z^{-4}\bigr)\biggr)
	\\
& \times \begin{cases} 0 , & \arg z \in [-\frac{3\pi}{4} + \delta, \frac{3\pi}{4} - \delta], \\
e^{i\pi a}, & \arg z \in [\frac{\pi}{4} + \delta, \frac{5\pi}{4} - \delta], \\
e^{-i\pi a}, & \arg z \in [-\frac{5\pi}{4} + \delta, -\frac{\pi}{4} - \delta], \\
\end{cases} \qquad z \to \infty, \  a \in \C,
\end{align*}
where the error terms are uniform with respect $a$ in compact subsets and $\arg z$ in the given ranges. 
It follows that
\begin{align*}
\psi_{11}(q,z) = &\; 
z^{i\nu} e^{-\frac{i z^2}{4}}\biggl(1 - \frac{i\nu(i\nu-1)}{2i z^2} + O\bigl(z^{-4}\bigr)\biggr)
	\\
& - \frac{\sqrt{2\pi}e^{\frac{i z^2}{4}}  z^{-i\nu-1} e^{\pi \nu}}{\Gamma(-i\nu)}  \biggl(1 + \frac{(i\nu+1)(i\nu+2)}{2iz^2} + O\bigl(z^{-4}\bigr)\biggr)  
	\\
&\times \begin{cases} 0 , &  \arg z \in [\delta, \pi] \cup [-\pi + \delta, 0], 
	\\
e^{-\frac{3\pi \nu}{2}}e^{\frac{3i\pi}{4}}, & \arg{z} \in [0, \frac{\pi}{2} - \delta], 
	\\
e^{\frac{\pi \nu}{2}}e^{-\frac{i\pi}{4}} , &  \arg{z} \in [-\pi, -\frac{\pi}{2} - \delta],
\end{cases} \qquad z \to \infty,
\end{align*}
and
\begin{align*}
\psi_{22}(q, z) = & 
z^{-i\nu} e^{\frac{iz^2}{4}}\biggl(1 + \frac{i\nu(i\nu+1)}{2iz^2} + O\bigl(z^{-4}\bigr)\biggr)
	\\
& - \frac{\sqrt{2\pi}e^{\frac{-iz^2}{4}}z^{i\nu-1}e^{\pi\nu}}{\Gamma(i\nu)} \biggl(1 + \frac{(-i\nu+1)(-i\nu+2)}{-2iz^2} + O\bigl(z^{-4}\bigr)\biggr)
	\\
& \times \begin{cases} 0 , & \arg z \in [0, \pi - \delta] \cup [-\pi, -\delta], 
	\\
 e^{\frac{\pi\nu}{2}}e^{\frac{i\pi}{4}}, & \arg z \in [\frac{\pi}{2} + \delta, \pi], 
 	\\
e^{-\frac{3\pi\nu}{2}}e^{-\frac{3i\pi}{4}}, & \arg z\in [-\frac{\pi}{2} + \delta, 0], 
\end{cases} \qquad z \to \infty,
\end{align*}
uniformly with respect to $a$ in compact subsets and $\arg z$ in the given ranges. 
Using the identity
$$\frac{d}{dz}D_a(z) = \frac{z}{2}D_a(z) - D_{a+1}(z)$$
we find similar asymptotic formulas for the derivatives of $\psi_{11}$ and $\psi_{22}$. 
Substituting these formulas into the defining equations (\ref{tildemcexplicit}) and (\ref{tildemcmcrelation}) for $m^X$, equation (\ref{mcasymptotics}) follows from a long but straightforward computation.
\proofendcontinue

It only remains to show that $m^X(q, \cdot) \in I + \dot{E}^2(\C \setminus X)$ and that $\sup_{z \in \C \setminus X} |m^X(q, z)|$ is bounded uniformly with respect to $q$ in compact subsets of $\D$. This is an easy consequence of Lemma \ref{EpCnlemma}, the asymptotics (\ref{mcasymptotics}), and the explicit formula (\ref{tildemcmcrelation}) for $m^X$. 
\proofend

\bigskip
\noindent
{\bf Acknowledgement} {\it The author is grateful to the referee for several valuable suggestions. Support is acknowledged from the EPSRC, UK, the Swedish Research Council Grant No. 2015-05430, the G\"oran Gustafsson Foundation, Sweden, and the European Research Council, Consolidator Grant No. 682537.}

\bibliographystyle{plain}
\bibliography{is}

\end{document}